%% file: mainArXiv2024.tex
\PassOptionsToPackage{table}{xcolor}
\documentclass[manuscript,nonacm,screen]{acmart}

\usepackage{graphicx}
\usepackage{hyperref}
\usepackage{mathtools}
\usepackage{caption}
\usepackage[gen]{eurosym}
\usepackage{color}

\input{preamble}

\AtBeginDocument{%
  \providecommand\BibTeX{{%
    \normalfont B\kern-0.5em{\scshape i\kern-0.25em b}\kern-0.8em\TeX}}}

\acmJournal{TOPS}

\begin{document}
\title{LightSC: The Making of a Usable Security Classification Tool for DevSecOps}

\author{Manish Shrestha}
\email{manish2sth@gmail.com}
\affiliation{%
  \institution{Department of Technology Systems, University of Oslo}
  \city{Kjeller}
\country{Norway}
}

\author{Christian Johansen}
\email{christian.johansen@ntnu.no}
\affiliation{%
  \institution{Department of Information Security and Communication Technologies, NTNU}
  \city{Gj\o{}vik}
\country{Norway}
}

\author{Johanna Johansen}
\email{johanna@johansenresearch.info}
\affiliation{%
  \institution{Department of Computer Science, NTNU}
  \city{Gj\o{}vik}
\country{Norway}
}

\renewcommand{\shortauthors}{Shrestha, et al.}

\begin{abstract}
DevSecOps is the extension of DevOps with security training and tools included throughout all the phases of the software development life cycle. DevOps has become a popular way of developing modern software, especially in the Internet of Things arena, due to its focus on rapid development, with short release cycles, involving the user/client very closely.
Security classification methods, on the other hand, are heavy and slow processes that require high expertise in security, the same as in other similar areas such as risk analysis or certification. 
As such, security classification methods are hardly compatible with the DevSecOps culture, which to the contrary, has moved away from the traditional style of penetration testing done only when the software product is in the final stages or already deployed.

In this work, we first propose five principles for a security classification to be \emph{DevOps-ready}, two of which will be the focus for the rest of the paper, namely to be tool-based and easy to use for non-security experts, such as ordinary developers or system architects.
We then exemplify how one can make a security classification methodology DevOps-ready. We do this through an interaction design process, where we create and evaluate the usability of a tool implementing the chosen methodology.
Since such work seems to be new within the usable security community, and even more so in the software development (DevOps) community,
we extract from our process a general, three-steps `recipe' that others can follow when making their own security methodologies DevOps-ready. 
The tool that we build is in itself a contribution of this process, as it can be independently used, extended, and/or integrated by developer teams into their DevSecOps tool-chains. Our tool is perceived (by the test subjects) as most useful in the design phase, but also during the testing phase where the security class would be one of the metrics used to evaluate the quality of their software.
\end{abstract}



\maketitle

\input{introduction}

\input{devsecops}
\input{methods}

\input{SC_process}

\input{tool_support}

\input{conclusion}

\paragraph*{Acknowledgements:}
We would like to thank Maunya Doroudi Moghadam and Josef Noll for help in different stages of this work.

\bibliographystyle{ACM-Reference-Format}
 \bibliography{database}

\input{appendix}

\end{document}

%% file: preamble.tex
\newcommand{\cj}[1]{}
\newcommand{\jj}[1]{}
\newcommand{\ms}[1]{}

\newcommand{\obselete}[1]{}
\newcommand{\TRonly}[1]{}

%% file: introduction.tex
\section{Introduction}
According to International Data Corporation, the predicted number of Internet of Things (IoT) devices for 2025 is 41.6 billion, generating ca.~7.9 zettabytes of data\footnote{\url{https://www.idc.com/getdoc.jsp?containerId=prUS45213219}}.
Because of this amount of produced data and human life penetration (e.g., in smart homes, offices, cities, hospitals), it is highly essential to develop secure IoT systems. However, securing IoT still proves challenging, especially in industries driven by functionality and low costs, demanded by the high competition in this new market, as argued, e.g., by~\cite{IoTSecurity15asiaCCS,khan2018iot,IoT19Security}.

IoT software, like most modern software, is developed in an agile style (see e.g.,~the Scrum\footnote{\url{https://ScrumGuides.org}} method), where popular now is the DevOps culture \cite{davis2016effective,humble2018accelerate,wiedemann2019research}. 
DevSecOps\footnote{\url{https://www.devsecops.org}} adds security tools and awareness at all phases of the software development life-cycle \cite{humble2010continuous}. However, security tools \cite[Part VI]{kim2016devops} need to have low learning and usability thresholds before they can be effectively included in the DevOps tool-chain \cite{farcic2016devops}.

Security is traditionally considered by the industry as an aftermath, a non-functional requirement that needs experts (e.g., white-hat penetration testing teams) to evaluate. Traditional methods like certification, risk analysis or security classification cannot keep up with the fast changing threat landscape in IoT systems \cite{nurse2017security}.
Standards such as ISO 27001 and certification such as Common Criteria are long and document-oriented processes. 
Keeping up with the software changes, in short and frequent release cycles as in agile, means updating the required documents regularly, which is not feasible. Similarly, labelling schemes such as UL Security Rating\footnote{\url{https://ims.ul.com/IoT-security-rating}} 
or BSI Kitemark\footnote{\url{https://www.bsigroup.com/en-GB/about-bsi/media-centre/press-releases/2018/may/bsi-launches-kitemark-for-internet-of-things-devices/}} are mostly based on penetration testing and risk analysis, besides documentation. 
Risk assessment methods 
(s.a.,
CORAS \cite{fredriksen2002coras}, EBIOS\footnote{\url{https://www.enisa.europa.eu/topics/threat-risk-management/risk-management/current-risk/risk-management-inventory/rm-ra-methods/m\_ebios.html}}, 
FAIR \cite{jones2004factor}, and OCTAVE \cite{alberts2003introduction})
require significant amounts of time and resources to conduct.  
These approaches follow a waterfall model where the assessments are far less frequent than the releases, and thus cannot fit the agile style of system development \cite{franqueira2011towards}.

As such, the software industry (and especially the IoT one) lacks motivation 
and guidelines for building security by design. We think that DevSecOps is one positive drive in this respect since it aims to lower the threshold for security aspects (e.g., tools, procedures, methods, guides) to enter the development process.

Security classification methods are not easy to integrate into the DevSecOps, and even more so for IoT \cite{bojanova2017trusting} where regulations, guidelines, and frameworks have only recently started to appear (see e.g., 
the Sancus architecture \cite{2017IoTsecFramework}, 
IoTSF\footnote{\url{https://www.iotsecurityfoundation.org/wp-content/uploads/2018/12/IoTSF-IoT-Security-Compliance-Framework-Release-2.0-December-2018.pdf}}, 
GSMA\footnote{\url{https://www.gsma.com/iot/iot-security-assessment/}},
IoT Working Group of 
CSA\footnote{\url{https://downloads.cloudsecurityalliance.org/assets/research/internet-of-things/future-proofing-the-connected-world.pdf}}, or the 
Industrial Internet Consortium\footnote{\url{https://www.iiconsortium.org/pdf/IIC\_PUB\_G4\_V1.00\_PB.pdf}}).

\paragraph{Contributions} Based on literature and our experience with security classifications and DevOps practices, 
we identify 
five principles for a security classification to be DevOps-ready.
In short, these are: (\ref{devsecops_req_dynamic}) dynamicity, (\ref{devsecops_req_tool}) tool-based, (\ref{devsecops_req_easy}) easy to use, (\ref{devsecops_req_impact}) static impact, and (\ref{devsecops_req_functionality}) oriented on protection mechanisms (detailed in Sec.~\ref{subsec_devsecops_req}).
We then choose an existing security classification methodology that already satisfies (\ref{devsecops_req_impact}) and (\ref{devsecops_req_functionality}) from \cite{shrestha2020methodology}, and focus here on making it satisfy the principles (\ref{devsecops_req_tool}) and (\ref{devsecops_req_easy}). Since the first principle is dependent on (\ref{devsecops_req_tool}), we discuss it as future work.

We are thus developing a tool, implementing the chosen methodology, and testing its usability on users selected to represent well our target group, i.e., non-security experts such as software developers, designers, architects, IT managers, or personnel from software operations.
Our users, described more thoroughly in Sec.~\ref{sec_methods}, are: 
(i) partners from one large European IoT project and students from one course on IoT security, both of which we involve several times during several stages of our development; 
as well as 
(ii) SMEs from a Polish cluster, and 
(iii) several developers recruited from software developing companies, 
both groups involved only for evaluating our high-fidelity web-based prototype.
To evaluate our prototypes and to extract information from our users, we organized workshops during which usability studies were run, involving methods such as interviews, observations, co-design, and active intervention, as well as standard questionnaires and recordings of user actions.

We do our work in five stages, developing three prototypes along the way; this is what we describe in Sec.~\ref{sec_manual_processes} (the manual stages) and Sec.~\ref{sec_tools} (the tool prototypes). In the end, we extract from this process a ``recipe'' to make it easy for others to transform security classifications (as well as other similar methods) into DevOps-ready tools, by following and maybe adapting our stages and instruments. We strive to make these stages intuitive and natural, following interaction design principles, but applied to our particular task of taking a complex, expert-oriented, method and transforming it into a tool that can be used by not-so-experts.
In short, one first needs to evaluate (see Sec.~\ref{subsec_SCM_eval}) the chosen security methodology as it is described in available documents or by experts; in our case, the methodology also has examples of applications to SHEMS (Smart Home Energy Management Systems) \cite{shrestha2019shems} and AMI (Advanced Meeting Infrastructure) \cite{shrestha2020methodology}.\footnote{Even if the methodology that we work with in this paper has been proposed by three of the current authors, we prefer to talk about the work in \cite{shrestha2020methodology} in third person, so to convey better to the reader that our process presented in this paper can be applicable to other methodologies. Moreover, the other two authors have functioned as 'newcomers' to this specific methodology, i.e., in tune with our target user group.}
Then one needs to transform the methodology into a process (steps to follow) focused on the non-expert target users (see Sec.~\ref{subsec_SCM_Process}).
The process then should be implemented into a low-fidelity prototype, e.g., in our case using spreadsheets (see Sec.~\ref{subsec_spreadsheet}), to test the automation and procedure nature of the method.
From the evaluation of the first implementation, one draws more concrete requirements for the high-fidelity version (see Sec.~\ref{subsec_web_v}).
In the end we evaluate (see Sec.~\ref{subsec_web_vv}) our final candidate for integration into a DevOps tool-chain.

Our current and future work is to help the software company eSmart Systems AS (which provides cloud-based solutions for smart grid monitoring of AMI) take up into their development process the tool that we present in this paper. From this point on we do not see significant research challenges, but only technical integration and maybe more iterations of UX adjustments/improvements to fit their specific development process and to enable dinamicity.

%% file: devsecops.tex
\section{Security Classification for DevSecOps}\label{sec_devsecops}

\subsection{DevSecOps and Usability of Security}

Agile methods \cite{cockburn2006agile} have been a popular style of software development for quite a while, adopting from the spiral model \cite{boehm1988spiral} the cyclic development, revisiting the same phase multiple times, e.g., new or changed requirements may be dictated by the client or the market. 
Agile promotes the inclusion of users, e.g., their manifesto\footnote{http://agilemanifesto.org/principles.html} encourages a software development culture that values:
(i)
 individuals and interactions
over processes and tools;
(ii)
 working software
over comprehensive documentation;
(iii)
 customer collaboration
over contract negotiation;
(iv)
 responding to change
over following a plan.

DevOps is a more recent agile style that differentiates itself through being open to and encouraging the use of tools at all phases, including the operations phase (thus the `Ops' in the name). Operations have become more important lately, both because of the proliferation of the cloud, making the infrastructure cheaper to deploy and run the software, and because of automation and tools becoming available for more tasks in all the development phases. 
DevSecOps brings into the DevOps the security, following the same philosophy, i.e., security awareness (or best practices) and security tools/processes at all phases. In particular, the penetration testing that depends on a high level of security expertise (usually coming from outside the development team) is mostly replaced by security tools such as code scanners, loggers, or API security testing, and phase relevant security education for all team members.

We consider DevSecOps as an arena that, more than ever, promotes the industrial adoption of usable security tools \cite{Cranor2005Security,karat2012privacy}.
On one hand, since DevSecOps is so tool intensive it lowers the usability threshold to allow more tools to be incorporated into the development tool-chain. 
On the other hand, since DevSecOps is so receptive to new tools, it offers researchers a motivation to put more effort into making their security tools easier to use, in the hope of being adopted by the industry.

\subsection{Principles for DevOps-ready Security Classifications}\label{subsec_devsecops_req}

We wish to propose five general \emph{principles for making a security classification DevOps-ready}, by which we mean a security classification that can be easily integrated into a DevSecOps tool-chain as one of the security mechanisms/tools.
These principles may be seen by some as also applicable to other similar expertise-heavy methods such as risk analysis, which are usually manual, slow, and expensive \cite{anderson2009certification}, requiring complex thinking to properly deal with uncertainties \cite{2020RiskThinking}.

If a reader not acquainted with security classifications may have difficulties following some of the, rather succinct, arguments behind the five principles, we hope that after going through the details of Sec.~\ref{security_classification}, the ideas presented below will be easier to appreciate. For now, we are contented with giving a brief definition of how we see a security classification to be (in very general terms).

\begin{quote}
A \emph{Security Classification Methodology} (SCM) has the goal to evaluate the security of a system with the outcome of classifying it; a security class offering a measure of the strength of the system. SCM (s.a.\ the ones from the French agency ANSSI or the US agency NIST) are often used for governmental systems, whereas similar methods for risk assessment (s.a.\ the standard ISO/IEC 27005 or the EBIOS from the European agency ENISA) are more often used by industry, and involve more calculations of losses and countermeasures in case of breaches. SCM compute a \emph{security class} by combining evaluations for: \emph{Impact} and \emph{Likelihood} (that the system is breached), where the likelihood is the result of combining the evaluations of the \emph{Exposure}, the users' \emph{Accessibility} to the system, and the power of \emph{Attackers}. Exposure in turn is determined by combining the \emph{Connectivity} and the security \emph{Protection} mechanisms supported by the system. (See also Fig.~\ref{fig:security_class_step} on page \pageref{fig:security_class_step}.)
\end{quote}

Based on our experiences with security classifications and with DevOps development practices, we consider the following principles as a minimum for a DevOps team to be able to adopt a new security classification methodology.

\begin{enumerate}
\item\label{devsecops_req_dynamic} \textbf{Dynamic.} 
In evergreen\footnote{https://www.danielengberg.com/what-is-evergreen-it-approach/} applications (e.g., nowadays web browsers\footnote{https://www.techopedia.com/definition/31094/evergreen-browser}) the development never ends, and updates (both functional and security/bugs patches) are constantly pushed to the deployed system, preferably without user interaction (e.g., no consent). Therefore, any security classification needs to be dynamic so that for each update, quick and cheap re-evaluations can be done  -- similar to how software testing is being done -- to cope with the short development life-cycles of DevOps.

\item\label{devsecops_req_tool} \textbf{Tool-based.} 
The method must have a tool support, and not only with a GUI but also with an API available, so that is can be integrated within the overall DevSecOps tool-chain (e.g., \cite{hsu2018hands}). Tools that are built with a UI (e.g., web-based apps) are also built with an API (e.g., RESTful) to which the UI connects, thus, an API is most often a byproduct of the tool development.

\item\label{devsecops_req_easy} \textbf{Easy to use for non-security experts.}
One of the main goals of DevSecOps is to move away from the traditional style of white-hat penetration teams who evaluate the security of a ready-built (often already deployed) system, and into a new style where every member of the DevOps team needs to have security competence relevant for their phase of development. Thus, a security classification method for DevSecOps needs to be usable by non-security experts.\footnote{The ``easy-to-use'' principle is related to, e.g., the addoption of privacy enhancing methodologies (PEMs) into software development processes where \cite{Senarath2019PEM} shows that complexity (i.e., the oposite of ease of use) has the most adverse impact (from all the five factors studied) on the developers' intention to use the respective PEM.} 

\item\label{devsecops_req_impact} \textbf{Impact statically and manually evaluated.}
Security classifications (the same as risk analysis methods) involve evaluating the impacts of security breaches. However, when using the security classification inside one company for developing one product, the impact evaluation is nearly static because the planned product and its functionalities and intended use, do not change radically during the lifetime of the product. Changes are usually very controlled, and those that are relevant for the evaluation of impact are even less frequent.
As such, the security methodology is enough to evaluate impacts once, in the beginning (maybe using even security experts), and input this evaluation manually into the tool. Therefore, we assume that impacts are of little concern for a DevOps-ready tool, and one need not spend effort on automating that.

\item\label{devsecops_req_functionality} \textbf{Fine-grained security functionality.}
Outside impact, security classifications are usually attack-centric, focusing on the capabilities of the attackers. For IoT and for a DevOps style of development, one would focus less on attackers, which are very dynamic and difficult to evaluate, and more on the security protection functionalities and exposures of the system under development, which are under the full control of the DevOps team. Focusing on functionalities makes it easy to automatically evaluate the system within a DevOps testing cycle, and also allows the developers to understand how to make their systems secure by design, by indicating which functionalities are a good match for which exposures and with what protection level (derived from the security class specifications).
\end{enumerate}

The methodology that we will work with is already developed to meet requirements \ref{devsecops_req_impact} and \ref{devsecops_req_functionality}. Thus we do not evaluate these here. Moreover, the dynamicity requirement can be achieved and evaluated only after a tool is built (see Sec.~\ref{sec_discussions}).
Therefore, in this paper, we focus on the two principles \ref{devsecops_req_tool} and \ref{devsecops_req_easy}.

%% file: methods.tex
\section{Participants}\label{sec_methods}\label{subsec_users}

The research in the paper has a user-centered approach, where the users and their goals are the driving force behind the development of a Security Classification Tool (SCT). 
Usability testing \cite{dumas1999practical} helps us discover problems with the chosen SCM and to develop an easy to use SCT for non-security experts.

\emph{Our target group} is \emph{non-security experts}, motivated by Principle~\ref{devsecops_req_easy}. More precisely, we are interested in people that have technology expertise, 
as well as people, such as system designers and developers, who are not security engineers but who may have basic security training (since their routine tasks need this) specific for their particular area of expertise. We are also interested in non-technology experts, like CEOs and managers of various development and operations aspects of technology; these people would know about use-cases, features, or economy and impacts, related to the technology system, but  not necessarily technical details.

\vspace{1ex}
\emph{The participants} involved in testing our prototypes are:
\begin{description}
\item[SCOTT project.] The most inputs and interactions were done with the participants from one large project called Secure Connected Trustable Things\footnote{https://scottproject.eu} (SCOTT) with ``57 partners from industry and academia from 12 countries working on 15 pilots involving 48 technological building blocks''. The main companies that we interacted with were: 
Philips Research\footnote{\href{https://www.philips.com/a-w/research/home}{https://www.philips.com/a-w/research/home}} (NL), 
Vemco\footnote{\href{https://vemco.pl/}{https://vemco.pl/}} (PL), 
AVL\footnote{\href{https://www.avl.com}{https://www.avl.com}} (AT), 
ISEP\footnote{\href{https://www.isep.ipp.pt}{https://www.isep.ipp.pt}} (PT),
VTT\footnote{\href{https://www.vttresearch.com/en}{https://www.vttresearch.com/en}} (FI)
and
Tellu IoT\footnote{\href{https://www.tellucloud.com/}{https://www.tellucloud.com/}} (NO),
as well as academics from 
Gdansk University of Technology\footnote{\href{https://eti.pg.edu.pl}{https://eti.pg.edu.pl}} (PL) and KTH (SE).

\item[Students.] These were attending one course on IoT security. There were relatively few student participants, but their inputs were valuable and representative for their target group (i.e.,the novice users). 

\item[SME cluster.] Through organizing a `hackathon' we reached out to a cluster of SMEs (Small and Medium-sized Enterprises) from Poland doing technology development.

\item[Software experts.] We also reached out to four individuals from industry who had long software development experience:
\begin{itemize}
\item Participant 1: CEO of a startup company with more than 25 years of experience in the software industry, especially on software used in the energy sector. His experience includes management and training, software design, development, and testing.
\item Participant 2: CTO of another company with more than 20 years of experience in the software industry, also having a good background in information security. 
\item Participant 3: Senior Consultant and Business Developer in another company with more than 20 years of experience in software development.
\item Participant 4: Software engineer with ca.\ 7 years of experience, having worked as a software engineer and data scientist in several companies.
\end{itemize}

\end{description}

In particular the SCOTT project participants were usually teams made of both technical and management people, and on rare occasions a person with considerable security expertise.
The `Software experts' category is, similarly, made of high-expertise people.
Rather to the contrary, the `Students' are technical people with little knowledge of security and fresh in the development field too.
The `SME cluster' was chosen so we can have teams that are more diverse in expertise, from business experts to developers (detailed in Sec.~\ref{subsubsec_hackathon}).

In our studies we were interested in testing with both individual users working alone (i.e., the `Students' and `Software experts'), but also with teams where the members collaborate in using the SC tool (i.e., the SCOTT and the SME cluster participants). Since our aim is to provide a SC tool for the DevSecOps team, both team work and individuals are important, as well as diversity of background, e.g., spanning the design, development, as well as the operations phases of DevOps. Our hackathon from Section~\ref{subsubsec_hackathon} is especially focused on diversity, whereas involving the individual `Software experts' in Section~\ref{subsubsec_individualTest_FinalTool} is meant to reach various types of DevOps work.

The users have been consulted throughout the development, and we explain in the rest of the paper how and for which of our studies we interacted with the different users from above.

%% file: SC_process.tex
\section{Manual Security Classification}\label{sec_manual_processes}

\input{background_on_SecClas}

\subsection{Evaluating the SC Methodology with Users}\label{subsec_SCM_eval}

The development of a Security Classification Tool (SCT) involved multiple stages of prototyping and usability testing.

\vspace{1ex}
\emph{The goal} of the first stage is to take the methodology as described in the research papers \cite{shrestha2020methodology,shrestha2019shems,shrestha2020building} and evaluate whether it follows the Principle~\ref{devsecops_req_easy}, i.e., that the SCM is easy-to-use for non-experts in security.

\vspace{1ex}
\emph{The Participants} in this evaluation stage were from two of our user groups, namely the Students and partners from the SCOTT project who were a mixture of technology people, with management and software/system design people; however, there were no security experts in their teams, except for some of the technology people who had general security knowledge or specific for their technical field.

\vspace{1ex}
\emph{Performing the test:} 
Our research team, which included security experts, first read relevant papers and understood from \cite{shrestha2020methodology} the SCM. We then prepared one presentation for the two groups of users.
(A) To the SCOTT partners, we presented and explained the SCM through a one hour workshop. The feedback was collected through structured conversations during a session after the presentation.
(B) To the students, we presented the SCM in one of the lectures 
and gave as a homework the research papers, which they were supposed to apply to their IoT system exercise%
and report back to the lecturer (one of our research team members).

\vspace{1ex}
\emph{The first results} were that none of the participants could understand the SCM, let alone how to apply it to their use cases. However, they did express interest in the concept of security classes. 
We did not obtain more concrete suggestions, mainly because the participants could not understand enough about SCM to give us meaningful comments.

Our team took then a second attempt at simplifying the presentation, and more importantly, we now presented how the SCM would be applied, focusing on exemplifying the work published in \cite{shrestha2019shems}.
We reasoned that by presenting an application of SCM to a similar IoT system, the participants would easily understand how to apply the SCM to their use case.
We also read various SCOTT project documents where their respective IoT systems were being described.
We then tried in our presentation to make, rather superficial, correlations between the application of the SCM from \cite{shrestha2019shems} and the participants' respective pilot systems. 
This second workshop with SCOTT did not manage to clarify enough as to allow the participants to apply the SCM. 
However, we did get more feedback during the structured conversations session. 
The topics included details of the SCM, like the calculation of impact and the evaluation of connectivity.

\vspace{1ex}
\emph{The final result} can be summarized, based on one of the participants observations, endorsed rather unanimously, as 
\begin{quote}
``It is not clear where to start with this methodology''. 
\end{quote}  

\emph{Explanations and Recommendations:} 
When reflecting on this observation, we could correlate it with how certification bodies use certification processes to do their work. An elementary definition of `process' implies a sequence of steps to be followed to arrive at a desired outcome. 
Having a predefined process for users to follow resonates with the external cognition approach \cite{scaife1996external}. 
Externalizing to reduce cognitive load means, in our case, producing a sequence of steps that a non-security expert could follow in order to evaluate the security class that a system belongs to.
Following the cognitive tracing technique, we decided to create a step-by-step process, meant to organize and externalize the requirements of the methodology and guide the users through the actions needed to perform a classification.

\subsection{SC methodology as a ten-step process}\label{subsec_SCM_Process}

We have structured the security classification methodology as a ten-steps process as follows:

\begin{enumerate}
\item\label{SCMstep1} \textbf{Define the IoT system.}
The user decides which system should be evaluated and gathers knowledge about the system, e.g.: system architecture, functionalities, security requirements, use cases, and context of use. This step helps the user to understand and prepare the system under evaluation.

\item\label{SCMstep2} \textbf{Define the components of the system.}
The necessary components of the system are defined, e.g., for a smart home one can have: IoT hub, smart devices, sensors, control data, etc.

\item\label{SCMstep3} \textbf{Describe the features of the components.}
The interactions between the system components are described. 
By now the user should have a reference architecture of the system and have identified a use case.

\item\label{SCMstep4} \textbf{Define the impact level.}
For each component, the worst impact of security breaches is defined. 
The impact levels are defined by the SCM research papers (following ANSSI) and is similar to the evaluation of impact in risk assessments. The impact may be on economy, human life, physical infrastructure, business, etc. 

\item\label{SCMstep5} \textbf{Describe communication mechanisms.}
The communication capabilities for each component are described, looking into which communication standards are used, e.g., WiFi, Bluetooth, LoRa6.

\item\label{SCMstep6} \textbf{Describe the type of networking.}
The user has to find out whether the network is only a Home Area Network or a Wide Area Network.

\item\label{SCMstep7} \textbf{Determine the connectivity level.}
Based on the two previous steps, the user assigns the connectivity level to the components. The connectivity level varies from C1 to C5 as described by the SCM.

\item\label{SCMstep8} \textbf{Determine the protection level.}
The user identifies relevant protection criteria for each component together with the respective security functionalities.
These are compared to the Protection Level table given by the SCM (see also Fig.~\ref{fig:excel_tool_support} on page~\pageref{fig:excel_tool_support}).

\item\label{SCMstep9} \textbf{Determine the exposure level.}
Use the information from the previous two steps in the lookup Table~\ref{table:exposure_lookup}.

\item\label{SCMstep10} \textbf{Determine the security class.}
Using the exposure and impact levels apply the lookup Table~\ref{table:classe_lookup}.
\end{enumerate}

Working with the SC methodology is manual, as far as the research papers \cite{shrestha2020methodology,shrestha2020building} describe it. Therefore, the above process is also manual, with the advantage 
that a clear procedure is given to the user to follow.
One can easily see that some of the above steps can be more or less automated. Automation is a highly desired method 
for
making a difficult technical process more user friendly, 
as it reduces the number of tasks the user 
has
to do.
Steps \ref{SCMstep1} to \ref{SCMstep3} are manual, and the user can take as much time and space for writing down the descriptions as required (no page limits).
Step \ref{SCMstep4} is a classical risk analysis stage, which we assume to be more static for DevOps and IoT software systems. This is also manual and requires security expertise.
Step \ref{SCMstep5} and \ref{SCMstep6} are also manual and needed only to help in step \ref{SCMstep7}.
Step \ref{SCMstep8} is probably the most tedious because of the long list of criteria that need to be evaluated. 
Steps \ref{SCMstep9} and \ref{SCMstep10} are mechanical tasks, done through lookup tables.

As such, steps \ref{SCMstep9} and \ref{SCMstep10} can easily be automated, whereas steps \ref{SCMstep1} to \ref{SCMstep7} not so easily; at least the SCM does not give us any help in that direction. Step \ref{SCMstep8} can be partly automated by summing up all the answers of the user and comparing them automatically with the respective table from the SCM.

\subsubsection{Evaluation of the ten-steps process}\label{subsec_SCM_Process_Eval}

Designing and evaluating the ten-step process was done over several workshops (each of 30min to 1h) interacting with the SCOTT users only. The major activity during this stage was to apply the SCM ten-steps to the pilots from SCOTT, together with the respective partners.

\vspace{1ex}
We had \emph{two goals}: 
\begin{enumerate}
\item The SCOTT users to understand how the SCM works and how to use it to apply it themselves.
\item Us to understand how easy it is to apply the ten-steps process to the IoT systems of the SCOTT pilots that we chose as test cases.
\end{enumerate}
For both goals, our interactions were geared towards collecting information about the usability of the ten-steps and how to improve it to fit the two examples that we considered representative of the intended application area.

\vspace{1ex}
\emph{The participants} were the two teams that were working on the two SCOTT pilots detailed below. During each workshop we had between two and four persons, 
where one was in management position (from the coordinating team of the respective pilot) and had broad knowledge about the respective system
and the others were technical people closely involved in the developing team (e.g., from GUT, Tellu IoT, AVL). 
These two teams of users have continued to interact with us until the last stage and the high-fidelity prototype.

\vspace{1ex}
\emph{We performed our studies} on two applications:
\begin{enumerate}
\item\label{subsec_SCM_Process_Eval_Philips} The ``Elderly UI'' component of the ``Assisted Living and Community Care System'' (ALCCS) pilot, coordinated by Philips Research. 
In short, the Elderly UI (see Fig.~\ref{fig_elderlyUI}) is a small form factor prototype device that can be worn as a patch on the skin 
for weeks at a time without the need for recharging, and is able to continuously observe activity and position from the elderly resident, and periodically transmits the observations straight to the Cloud. 

\begin{figure}[htb]
\centering
  \includegraphics[width=5cm]{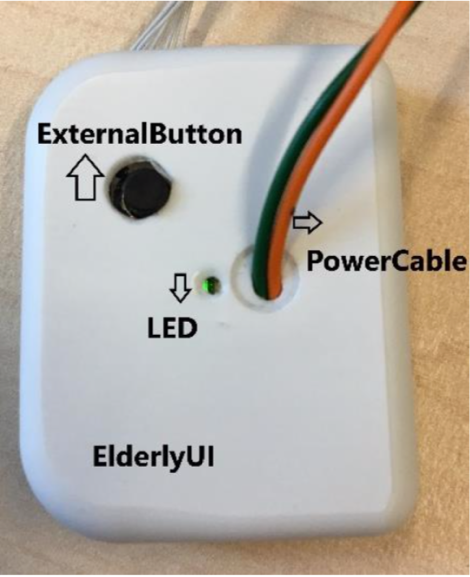}
  \caption{Early prototype of the ElderlyUI component. \\(Description and image, courtesy of Philips Research.)}
  \label{fig_elderlyUI}
\end{figure} 

\item The ``Multimodal Positioning System'' (MPS) component of the ``Secure Connected Facilities Management'' pilot, coordinated by Vemco.
The MPS had as main functionality the localisation of people and assets within critical infrastructures, being applied in this case inside a refinery.
\end{enumerate}

We ran two workshops for each test case. 
During these, the ten-steps process went through two major redesigns, where mainly the order and number of the steps were changed, and the helping descriptions were improved.

In each workshop we used the co-discovery technique \cite{kennedy1989using,lim1997empirical}, which is especially useful in such an early design phase, with discussions going between us, the research team, and the respective SCOTT team. Thus, during these workshops, we adjusted our understanding of the test systems and worked with the teams to understand how to properly apply the SCM to their systems.
As materials, besides the previous presentations, we also used the technical project-internal documents for each test system to collect the necessary information for evaluating the connectivity, protection, and exposure levels. 

Each workshop also employed the active intervention technique, which is excellent in discovering a wealth of diagnostic information about the prototype \cite{dumas1999practical}, which in our case was the ten-steps process.
We were guiding the SCOTT team, meaning us, e.g., 
explaining the purpose of a step (often mostly confirming that their understanding of that step was matching with ours); or
giving more details about a step like what was meant by the Home Area Network.

One playful activity that our users enjoyed was to work on identifying how the security class can be improved; e.g., for the ElderlyUI system we had scenarios that changed the class from E to B by making changes to the system. 
This is one major benefit claimed by the main article \cite{shrestha2020methodology} of the security classification methodology. Therefore, our interactions confirm this claim that IoT developers would enjoy knowing the security class of their system, which in turn would encourage them to strive to improve their system's security so to improve the class.

\subsubsection{Major findings}\label{subsec_SCM_Process_Findings}

Besides the constant feedback that we received during the workshops about small improvements to the ten-steps process, we made the following major observations.
\begin{enumerate}
\item The participants could now perform most of the ten steps, under our guidance. 
\item The most difficult parts of the methodology were identified as being:
\begin{enumerate}
\item\label{find_difficult_impact} the evaluation of the Impact level, which, they said: 
``Looks like the job of a security expert'' 
(which the participants were not);
and 
\item\label{find_difficult_protection} finding the Protection level, since it involved answering many specific security questions that needed interactions with other members of their development teams.
\end{enumerate}
\end{enumerate}

\vspace{1ex}
\emph{Explanations and Recommendations.}
Regarding the observation \ref{find_difficult_impact}, the SCM papers \cite{shrestha2020methodology,shrestha2019shems} especially point out that the evaluation of the impact level is not a specific concern of the SCM and is supposed to be similar to how risk assessment or similar methods evaluate impacts of attacks. Moreover, the impact level is only indicative and does not need to be done to a perfect detail for one to use the SCM as it was intended. 
Recall 
from the Introduction 
that our goal in this work is to take a security classification methodology as it is, and make it DevOps-ready by building a tool that makes it easy for non-security expert users to apply it.
Therefore, 
since we are taking the SCM as given, and do not aim to improve it as a security instrument per se,
we decided that based on observation \ref{find_difficult_impact} we would only improve in the future tool the help text for this step by explaining the above aspect to the users so that they can get over this step with less concern.

The second observation \ref{find_difficult_protection} is, however, directed to a core aspect of the chosen SCM, since the list of security functionalities that the observation refers to, is a main differentiating aspect claimed by \cite{shrestha2020methodology,shrestha2019shems}. Therefore, we decided to improve on how the users work with this list in the next iteration.

%% file: background_on_SecClas.tex
\subsection{Reviewing the Security Classification Methodology}\label{security_classification}

The security classification methodology that we take as the starting point in this work has been proposed in \cite{shrestha2020methodology} as an extension of the standard for ``Security Classification of Complex Systems'' developed by the French national agency ANSSI. Besides, the methodology of \cite{shrestha2020methodology} incorporates security concepts from (and conforms with) several other relevant standards, among others, ISO/IEC, ETSI, OWASP, ENISA. This method has been detailed and extended towards IoT systems in \cite{shrestha2020building}.
We give here a very short review of this specific SCM, since more details will appear in Sec.~\ref{subsec_SCM_Process}.

\begin{figure}[htb]
\centering
\vspace{-3ex}\includegraphics[width=\columnwidth]{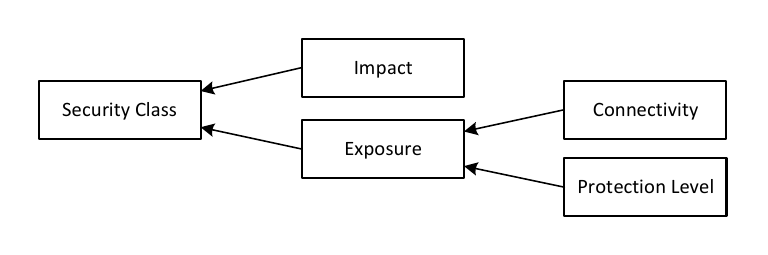}\vspace{-3ex}%
\caption{Components of the evaluation of a security class.}
\label{fig:security_class_step}
\end{figure}

The methodology is based on the analysis of impacts, connectivity, and protection level of the system. Protection level is determined from the protection mechanisms that are applied to the system. Protection level combined with connectivity forms the exposure level, and finally, exposure and impact are used to determine the security class of the system, as displayed in Fig.~\ref{fig:security_class_step}.
SCM considers five levels of Connectivity \cite[Sec.3.1]{shrestha2020methodology} adopted from ANSSI.

The protection mechanisms are evaluated based on a list of security criteria \cite[Table 3]{shrestha2019shems} that sum up to a protection level (from P1 to P5). The higher the protection level, the more security mechanisms it includes (when relevant, e.g., for the connectivity of the system). 
Finally, the classification methodology considers five impact levels -- also taken from ANSSI (see \cite[Sec.3.7]{shrestha2019shems}) -- namely Insignificant, Minor, Moderate, Major and Catastrophic. The impact level is determined usually by security experts (as mentioned by Principle~\ref{devsecops_req_impact} in Sec.~\ref{subsec_devsecops_req}).

A lookup table is used to determine the exposure from connectivity and protection levels as shown in Table \ref{table:exposure_lookup}. Finally, the security class is determined from the exposure and impact using a class lookup Table \ref{table:classe_lookup}.

\begin{table}[tbp]
\begin{minipage}{0.5\textwidth}
\centering
\caption{Calculations of Exposure Levels.}\label{table:exposure_lookup}
\begin{tabular}{|p{2.5cm}|c|c|c|c|c|}
\hline 
P1 & \cellcolor[HTML]{C0813E}E4 & \cellcolor[HTML]{C0813E}E4 &  \cellcolor[HTML]{9F2E4B}\color{white}E5 &  \cellcolor[HTML]{9F2E4B}\color{white}E5 &\cellcolor[HTML]{9F2E4B}\color{white}E5 \\ 
\hline 
P2 & \cellcolor[HTML]{EA8446}E3 & \cellcolor[HTML]{C0813E}E4 & \cellcolor[HTML]{C0813E}E4 &  \cellcolor[HTML]{9F2E4B}\color{white}E5 &\cellcolor[HTML]{9F2E4B}\color{white}E5  \\ 
\hline 
P3 & \cellcolor[HTML]{F6F166}E2 & \cellcolor[HTML]{EA8446}E3 & \cellcolor[HTML]{EA8446}E3 &  \cellcolor[HTML]{C0813E}E4  & \cellcolor[HTML]{C0813E}E4 \\ 
\hline 
P4 & \cellcolor[HTML]{40A377}E1 & \cellcolor[HTML]{40A377}E1 & \cellcolor[HTML]{F6F166}E2 &  \cellcolor[HTML]{F6F166}E2  & \cellcolor[HTML]{EA8446}E3 \\ 
\hline 
P5 & \cellcolor[HTML]{40A377}E1 & \cellcolor[HTML]{40A377}E1 & \cellcolor[HTML]{40A377}E1 &  \cellcolor[HTML]{40A377}E1  & \cellcolor[HTML]{F6F166}E2 \\ 
\hline 
\textbf{Protection/\newline Connectivity} & C1 & C2 & C3 & C4 & C5 \\ 
\hline 
\end{tabular} 
\end{minipage}
\hspace*{-\textwidth} \hfill
\begin{minipage}{0.5\textwidth}
\centering
\caption{Calculations of Security Classes.}\label{table:classe_lookup}
\begin{tabular}{|p{2.3cm}|c|c|c|c|c|}
\hline 
Catastrophic & \cellcolor[HTML]{40A07B}A & \cellcolor[HTML]{EA8446}C &  \cellcolor[HTML]{A32C47}\color{white}E & \cellcolor[HTML]{6B0000}\color{white}F &  \cellcolor[HTML]{6B0000}\color{white}F \\ 
\hline 
Major & \cellcolor[HTML]{40A07B}A & \cellcolor[HTML]{F6F166}B & \cellcolor[HTML]{C0813E}D &  \cellcolor[HTML]{A32C47}\color{white}E &  \cellcolor[HTML]{6B0000}\color{white}F \\  
\hline 
Moderate & \cellcolor[HTML]{40A07B}A & \cellcolor[HTML]{F6F166}B & \cellcolor[HTML]{EA8446}C & \cellcolor[HTML]{A32C47}\color{white}E &  \cellcolor[HTML]{A32C47}\color{white}E \\ 
\hline 
Minor & \cellcolor[HTML]{40A07B}A & \cellcolor[HTML]{40A07B}A & \cellcolor[HTML]{F6F166}B &  \cellcolor[HTML]{C0813E}D  & \cellcolor[HTML]{C0813E}D \\ 
\hline 
Insignificant & \cellcolor[HTML]{40A07B}A & \cellcolor[HTML]{40A07B}A & \cellcolor[HTML]{40A07B}A &  \cellcolor[HTML]{EA8446}C  &  \cellcolor[HTML]{EA8446}C \\ 
\hline 
\textbf{Impact/\newline Exposure} & E1 & E2 & E3 & E4 & E5  \\ 
\hline 
\end{tabular} 
\end{minipage}
\end{table}

%% file: tool_support.tex
\section{Creating the SC tool}\label{sec_tools}

In this section we describe how we built an online tool implementing the ten-steps process, and how we tested it with users during multiple testing sessions for different prototypes. 
To further consolidate the cognition support, we followed the computational offloading principle \cite{scaife1996external} and built a tool to help the user with their tasks by organizing, guiding, and automating some of the aspects of the task.
In our case, the task at hand was the process of security classification, which also had some of the steps ready for automation; whereas for the other steps the tool was intended to help with organizing the work and gather inputs from the users.

\subsection{Spreadsheet implementation}\label{subsec_spreadsheet}

Our first low fidelity prototype was in form of a spreadsheet and was implemented in Google Sheets because, as a cloud application, it allows a team to collaborate in real-time.

\begin{figure*}
\centering
  \includegraphics[width=\textwidth]{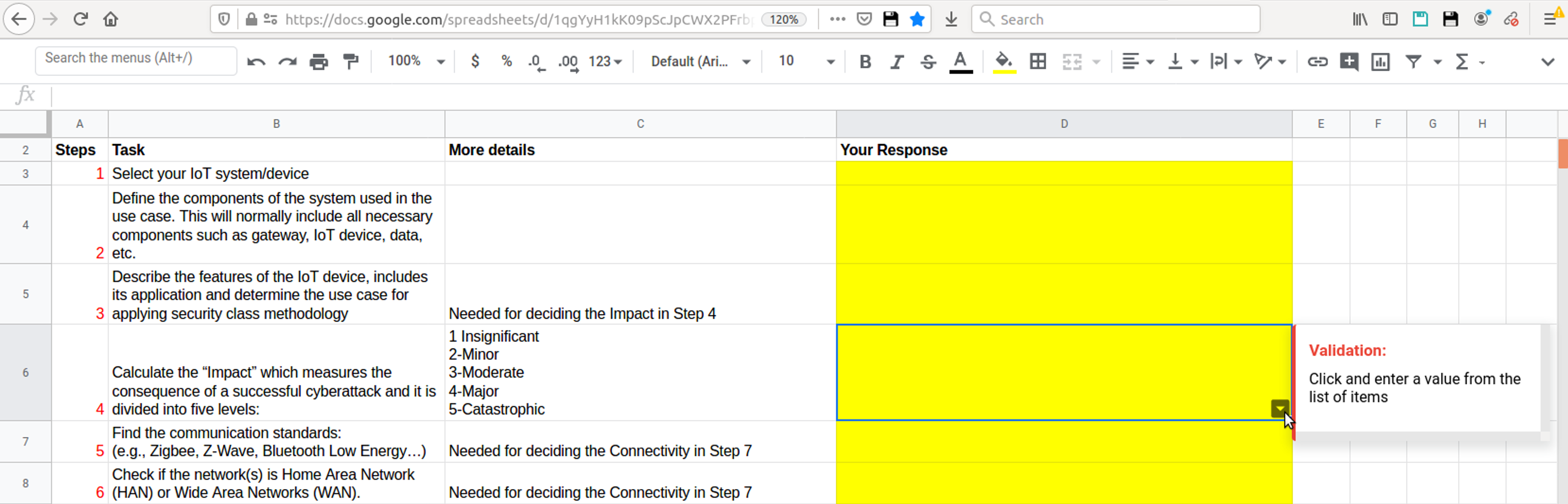}
  \includegraphics[width=\textwidth]{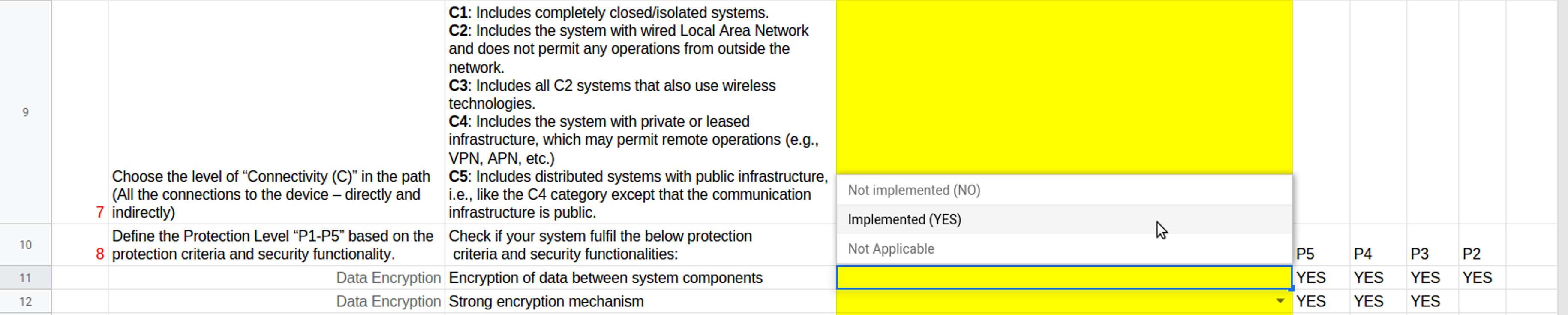}
  \includegraphics[width=\textwidth]{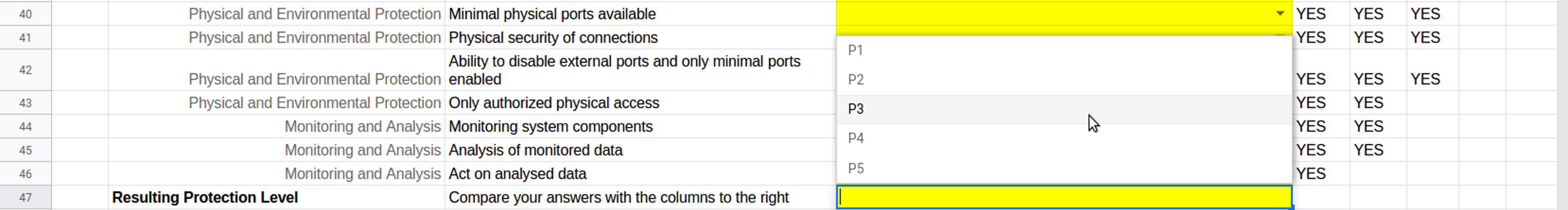}
  \includegraphics[width=\textwidth]{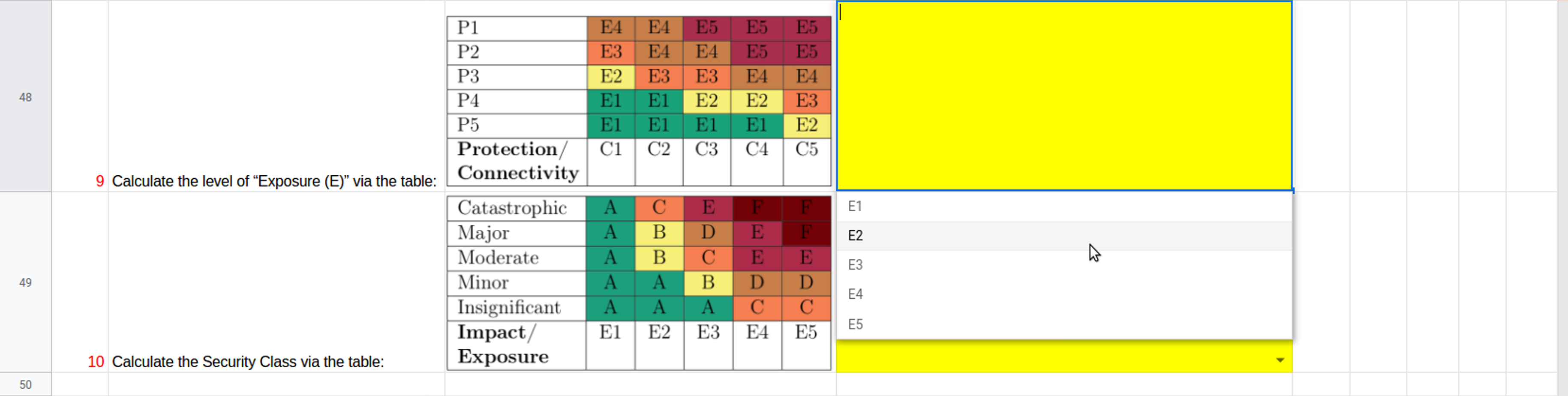}
  \includegraphics[width=\textwidth]{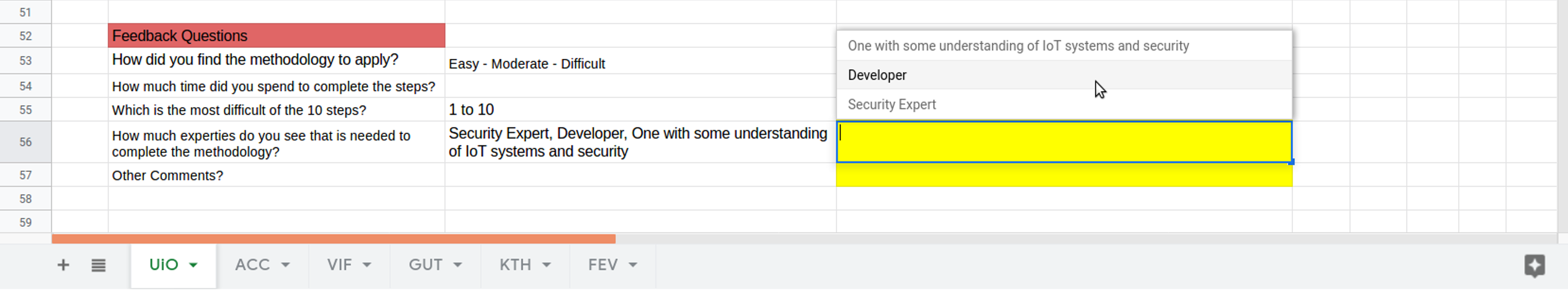}
  \caption{Snapshot of spreadsheet implementation of the SCM ten-steps process.}
  \label{fig:excel_tool_support}
\end{figure*}

The spreadsheet template (see Fig.~\ref{fig:excel_tool_support}) contained all the information from the previous ten-steps process, albeit in a more structured way, having the following components:
\begin{description}
\item[Step:] The step number coordinates the attention of the user and helps direct the workflow.
\item[Task:] A column providing the task description; the text is adopted from the ten-steps described before.
\item[More details:] This column provides additional descriptions to make the task easier to understand.
\item[Your Response:] This column stores the input from the user to the respective task, collected in three ways:
\item[-- Free Text:] The users could freely describe the system with components and relevant functionality.
\item[-- Dropdown list:] For inputs predefined by the methodology, requiring a specific item form a list (e.g., connectivity, protection level, presence of security functionality), we also applied a validation mechanisms to guide the users.
\item[-- Lookup table:] The respective lookup table was given for deciding the exposure and security class.
\item[Protection level requirements:] Additional information columns displayed protection level requirements, guiding the users to compare and select the appropriate protection level (see line 47 in Fig.~\ref{fig:excel_tool_support}).
\end{description}

\emph{Our goals} were  
to simplify the security classification process 
and
to present and test 
it %
with more users.
Therefore, for this low-fidelity prototype, we focused on providing clarifying text and necessary helper information for each step, based on the observations from Sec.~\ref{subsec_SCM_Process_Findings}.

\subsubsection{Performing the studies}

\emph{The participants} were SCOTT partners (seven teams, including the two from Sec.~\ref{subsec_SCM_Process_Eval}) and the same students from before, as detailed further.
The participants included security experts, developers and system managers having a general understanding of the IoT system and security.

\vspace{1ex}
\emph{One pilot test} was carried with AVL, one of the SCOTT partners. The result is the one presented in Fig.~\ref{fig:excel_tool_support} and is the one that we have used to do our final webinar tests.

\vspace{1ex}
\emph{We organised a webinar} for the whole SCOTT project partners, where the two teams from Sec.~\ref{subsec_SCM_Process_Eval} also helped us with the organization.
We used classical methods of advertising to attract many participants, like preparing an invitation text presenting the webinar (similar to how one would do for an academic event or hackathon, but more flashy) and e-mailing it to everyone in the project, with reminders, etc. We worked with the project coordinators to make the invitation text interesting for our audience, i.e., many of the SCOTT partners were companies.

The plan for the webinar was:
\begin{enumerate}
\item\label{webinar_presentation} An introduction from us, which included:
(i) a motivating presentation of the SC methodology, taken from the research papers,
(ii) a short presentation of the ten-steps process,
(iii) with an exemplification of how we used the ten-steps on the application from Sec.~\ref{subsec_SCM_Process_Eval}(\ref{subsec_SCM_Process_Eval_Philips}), meant as additional motivation and inspiration for the participants since it was from the same project.
\item\label{webinar_work} A hands-on interaction from the participants with the online spreadsheet.
\item\label{webinar_questions} A brief (since we were restricted by the time availability of our participants) questionnaire at the end of the spreadsheet (see bottom of Figure~\ref{fig:excel_tool_support}).
\end{enumerate}
The part (\ref{webinar_presentation}) took ca.~30min whereas parts (\ref{webinar_work}) and (\ref{webinar_questions}) some extra 30-40min, including final discussions.

We had ca.~15 participants in the online webinar (3 were the organisers). The participants were divided into five teams (based on the SCOTT pilot that they were working on) and took our hands-on exercise. Each team (see bottom of Fig.~\ref{fig:excel_tool_support}) had to fill in our spreadsheet template according to their IoT system of choice. The exercise took between 7-30min to complete. 

For part (\ref{webinar_work}) we used 
direct, unobtrusive
observation, where we were observing online how the teams were progressing. This was possible due to the capabilities of the Google Sheets to show the changes done by the participants, synchronously and in real-time.
At times we had to answer questions, usually for clarification or confirmation.

\vspace{1ex}
\emph{One final workshop} was done with the students, using a very similar setup and activity as above, during one hour of their exercise classes, i.e., we presented the spreadsheet tool and asked them to apply it to the same system as before, under our observation this time.

\subsubsection{Major findings}

From our observations and interactions during the webinar, we draw three conclusions.

\begin{description}
\item [User help/manual:]
Even if the spreadsheet and terminologies were explained in our presentation, all users still had questions either for clarifying individual steps or how to assign values for impact and connectivity. 

\item [Automation:] 
Several of the steps could be automated, e.g., determining the protection level, exposure, or class.
These were asked for by participants and supported by everyone.

\item [Lack of customisation:]
The spreadsheet did not allow to change the lookup tables, which participants observed as a necessity when changing the type of system. 

\end{description}

From the answers to our short questionnaire, we obtained the following:

\begin{description}
\item [Moderately difficult:]
All teams answered that they found the application of the methodology of \emph{moderate} difficulty.

\item [The difficult steps] 
were, again, the evaluation of \emph{impact} and the \emph{protection level} calculation.

\item [Diversity of expertise:]
Especially for answering all the questions for the protection level the teams needed diversity of expertise, i.e., they had to ask people that knew about the respective security functionality.
\end{description}

The student workshop confirmed that the ten-steps were now considerably easier to use than in the previous session when only the research papers were given.

\subsection{Web-based SC tool pilot testing}\label{subsec_web_v}

The high-fidelity SCT was implemented as a web application.\footnote{The final Security Classification Tool is available at \url{https://lightsc.azurewebsites.net}.}
The major technologies used were the following.
The development used ASP .NET Core and the Model View Controller pattern \cite{gamma1994design},
implementing also a separate service layer to provide a public RESTful API, useful when integrating in a DevOps tool-chain.
We used Azure SQL database for data persistence and deployed the application in Microsoft Azure cloud services.

\begin{figure*}[htb]
\centering
  \includegraphics[width=\textwidth]{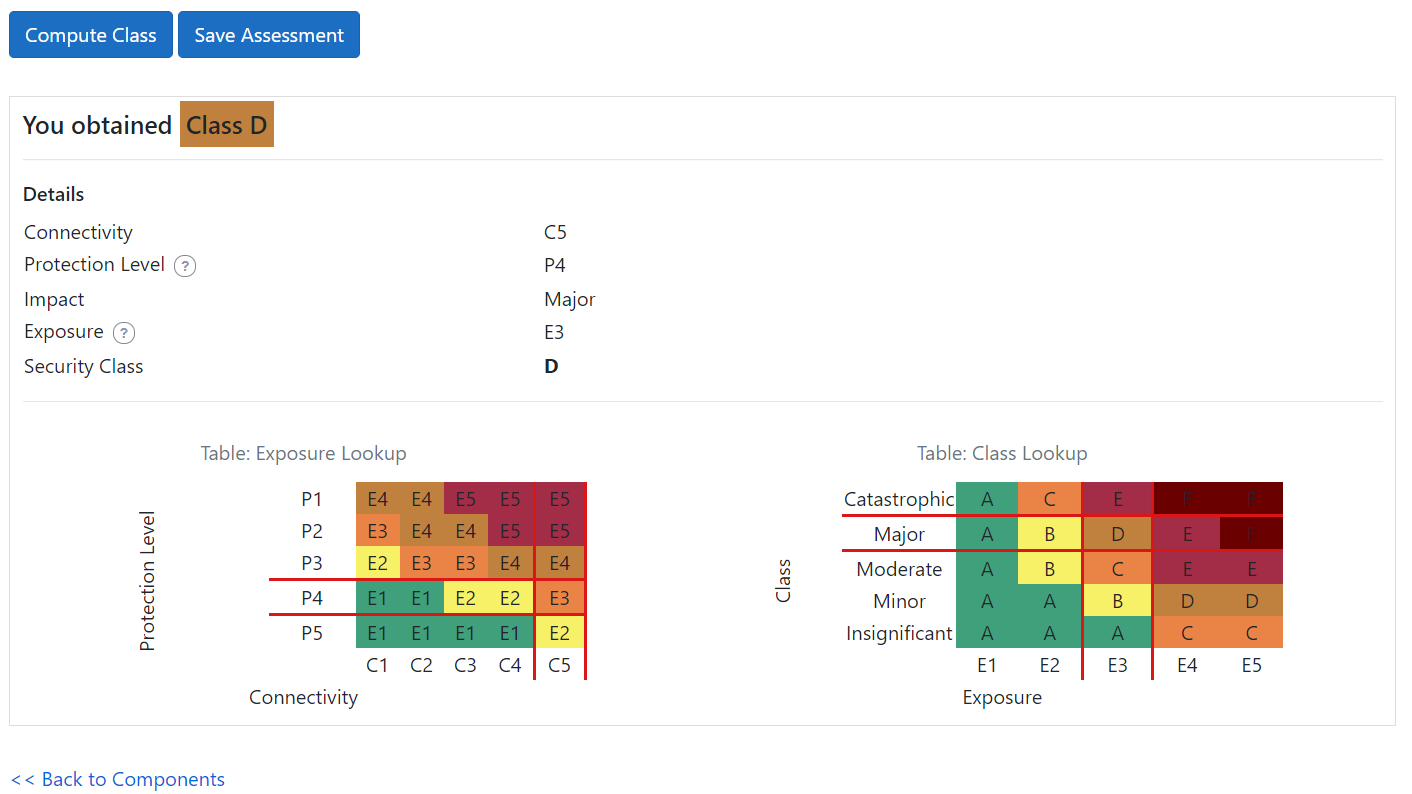}
  \caption{Snapshot of class calculation view.}
  \label{fig:compute_class}
\end{figure*}

We simplified the assessment process by combining several steps into one, with main activities now being:

\begin{enumerate}
\item \textbf{Define a System} (corresponding to step~\ref{SCMstep1} from Sec.~\ref{subsec_SCM_Process}) with a snapshot in Fig.\ref{fig:systems_page}.

\begin{figure*}[htb]
\centering
  \includegraphics[width=\textwidth]{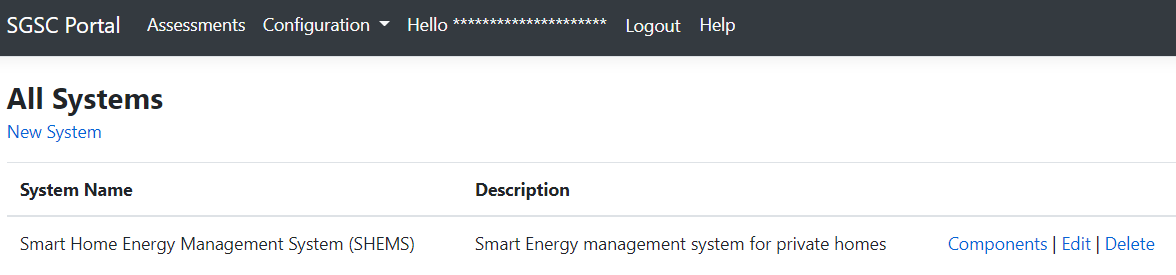}
  \caption{Snapshot of systems page of SCT web application.}
  \label{fig:systems_page}
\end{figure*} 

\item \textbf{Add components}  (implementing steps~\ref{SCMstep2}--\ref{SCMstep7} from Sec.~\ref{subsec_SCM_Process}).
A system is decomposed into its components, and for each component in turn a class can be computed.\footnote{See a video tutorial on the Help page of our tool: \url{https://lightsc.azurewebsites.net/UserHelp/VideoTutorial}} 
With the web tool we could add more organizational element, most importantly,
components can now be categorized, providing as default component types: IoT device, Hub, and Backend System. The user can define their own component types. The component types are relevant for the next step so that the tool can select automatically some of the security functionalities as `not applicable'.

\item \textbf{Perform assessment} (implementing step~\ref{SCMstep8} from Sec.~\ref{subsec_SCM_Process}) where security functionalities are selected.

\item \textbf{Compute security class} (automating steps~\ref{SCMstep9}--\ref{SCMstep10} from Sec.~\ref{subsec_SCM_Process}) 
by pressing a button.
Fig.~\ref{fig:compute_class} shows the final view containing also the lookup tables and what selections were made to obtain the resulting class.
\end{enumerate}

\subsubsection{Pilot testing}

The application was demonstrated 
to the SCOTT partners AVL and GUT, i.e.,
two of our main teams with whom we interacted several times during all this work. 
We had one workshop where we presented 
the new web application and demonstrated how to apply it to the original SHEMS example from research paper \cite{shrestha2019shems}.
After the presentation, and during the demonstration, we had a long period of discussions with comments from the users.
We did not perform full scale applications because these users already knew and had applied the ten-steps process.

The improvements have been appreciated, especially the save functionality and the login possibility since it allowed for a private space for someone to work with their evaluations. The automation was as expected.

The negative comments were especially related to the lack of help and guidance. One specific request was to have tool-tips for various parts of the interface, to give them local information (the screenshots in this paper are taken from the final version where this feature was implemented).

\subsection{Final version of SCT}\label{subsec_web_vv}

The final version of the web application had the following extra usability functionalities:
\begin{enumerate}
\item \textbf{Customisable lookup tables.}
Lookup tables are usually constructed by experts. The default ones that the application offers are the ones we took from the research papers of the SCM \cite{shrestha2020methodology,shrestha2019shems}.
However, as we learned from the users, depending on the domain of application, the lookup table may differ slightly. Therefore, one should be able to change the lookup table according to their domain. The tool has a configuration feature where the user can override the default lookup table and also reset it to default. 
\item \textbf{Main user guide easily available on every page.}
The preliminary tool had a user guide only on the landing page. Every time the user needed help, they had to browse to that page, which was considered hectic. 
This version introduces easily, and at all times, available help menu, now being placed as a sidebar which on click, slides over the page (see Fig.~\ref{fig:user_help_snapshot}). This sidebar allows the user to focus on their tasks, without the distraction of opening a new page each time help is needed. 

\begin{figure*}[htb]
\centering
  \includegraphics[width=\textwidth]{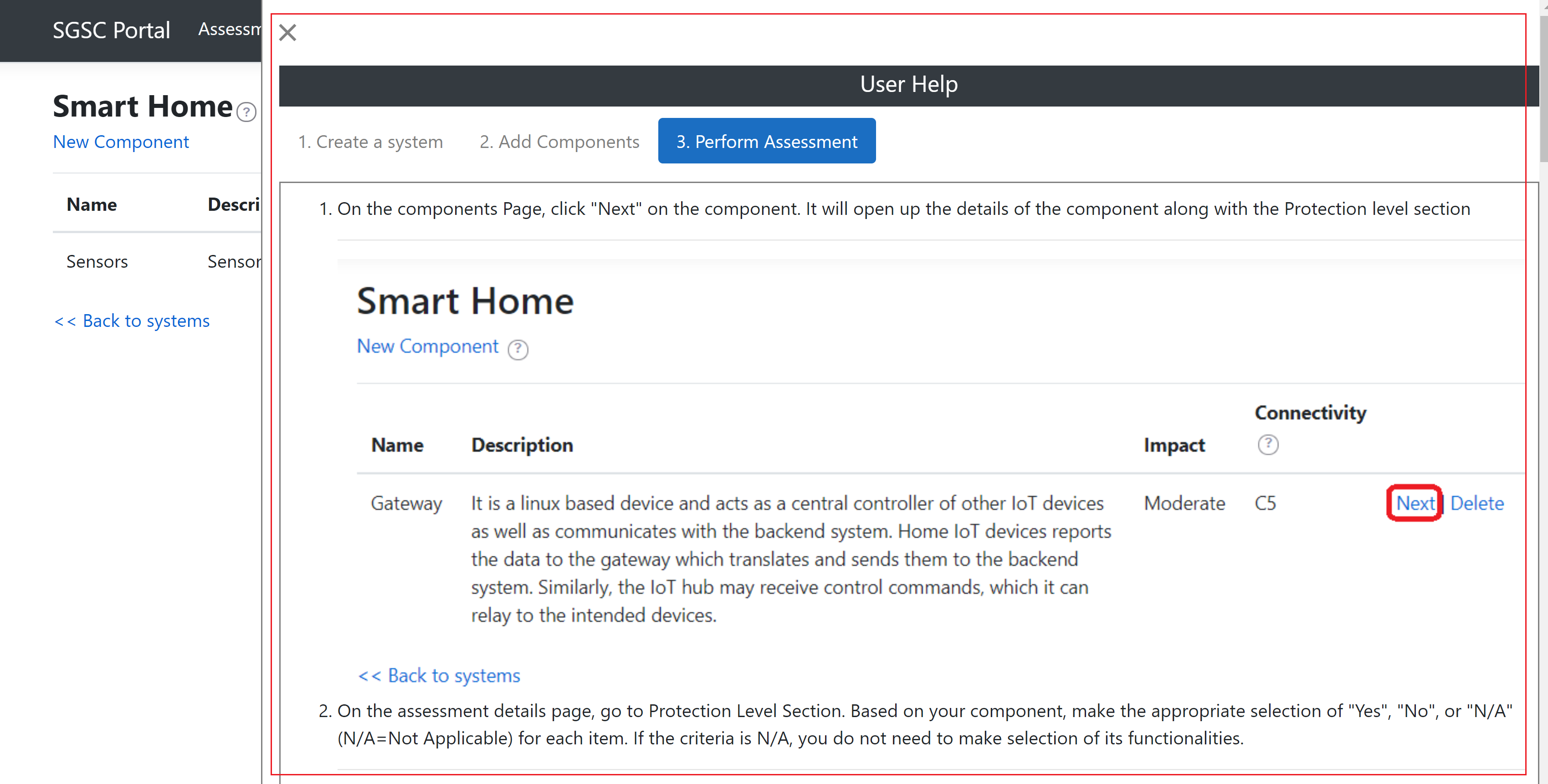}
  \caption{Snapshot showing user help opened in a side bar from the right.}
  \label{fig:user_help_snapshot}
\end{figure*} 

\item \textbf{Detailed contextual help.}
Since the users have constantly been asking for explanations of the terminologies and of the steps, 
we added help icons beside the respective texts or UI elements that required detailed explanations. When clicking on the help icon a modal window opens up to show these details. Many of these details also appear in the main help.
\end{enumerate}

In this version, we also decided to implement the \emph{``beliefs and weights''} aspect of the SC methodology from the research paper \cite{shrestha2020building}. Before, in the spreadsheet, it was difficult to work with these confidence parameters; but now the tool could more easily calculate using weights and the formulas from \cite{shrestha2020building}, with the user only specifying the individual weights.

\subsubsection{Evaluation through a Hackathon}\label{subsubsec_hackathon}

Helped by the SCOTT partner GUT (Gdansk University of Technology) we organised a hackathon contest with a cluster of Polish SMEs.

The \emph{preparations} for the hackathon included: 
(1) a video tutorial (ca.~10min) on how to use the tool;
(2) preparing a presentation with slides 
(a) motivating the concept of security classification, 
(b) describing the benefits for industry, 
(c) explaining the ten-steps process, and 
(d) how to apply it to the SCOTT pilot (this we mostly reused from previous workshops with additions and adaptations to fit the target audience); as well as
(3) materials for announcing and attracting participants and for managing the contest.

\emph{The hackathon event} had a ca.~one hour program (all was recorded through the online meeting tool) with: 
\begin{enumerate}
\item a short introduction (2min) from the SCOTT official and the Polish cluster official (our contact point), 
\item followed by our presentation and demonstration of the web tool, 
\item ending with the presentation of the contest, rules, tasks, and prizes (described further).
\end{enumerate}

\emph{The hackathon format} included a contest with three prizes (winning 2000\euro{} in total) and rules for participation and evaluation.
The \emph{contest} asked the teams to 
(1) use the tool on one of their systems or components; and
(2) describe how the security classes could contribute to innovation and business potential for their company.
Our purpose with preparing such a complicated setup was firstly to attract diversity in the participating teams, as well as hoping to increase the number of participants from industry. The contest was thus only a framing, where our real interest resided in the 
\emph{usability part} of the hackathon:

\begin{enumerate}
\item We offered special recognition prizes (with extra winnings and the title of ``Usability Wizards'') for those that take substantial effort to help us with the usability studies, i.e., to use the two aspects mentioned below. 

\item \emph{We prepared a survey} and asked the participants to take part in the survey, which was available through a special menu in the web interface.
The survey included questions regarding user experience, opinion about the tool, facts about the users, their expertise and knowledge of DevOps, and further suggestions.

\item We used Hotjar\footnote{https://www.hotjar.com/} to track and analyze users' activities (i.e., interaction logs) while they were performing their evaluations with our SCT. This method of indirect observation was necessary because our participants needed the flexibility regarding doing the task that the contest asked for.
We used the following particular strategies, detailed in Sec.~\ref{subsubsec_finalHotjar}:
(1) \emph{Screen recordings}
of the activity of the user while working with the web tool, captured anonymously for privacy concerns.;
(2) \emph{Incoming feedbacks},
with which the users could select the specific part of the page and provide feedback on it;
(3) \emph{Heatmaps}
 showing which part of the page was clicked, scrolled or hovered over the most. 
Using this method, we were able to identify which features the users are most interested in or are most difficult and require most effort/time. 
\end{enumerate}

\emph{The participants} attending the hackathon presentations were from four companies, of which three teams submitted the required report, with one team taking also the survey. 
The participation was poorer than we had expected, which was later explained by our local contact as \emph{``Language barrier''}, i.e., the writing in English was discouraging, and the internationalisation that the hackathon offered was not of interest since many of the cluster companies already had a large client base in Poland.
From the three reports that we have receive, one applied the SCM to a Mini Unmanned Surface Vessel, and they used the SCT to compare between a not secured version, that resulted in class F, and a secured version which resulted in class B. They claimed that this helped them understand what security functionalities the system needed. 
The other two applications were to analyze the 
security of autonomous vehicle management systems in logistics
and of RFID. Both reports used the tool similarly for trying out different security features for different configurations of their systems resulting in different security classes.

\subsubsection{Evaluation with Individuals}\label{subsubsec_individualTest_FinalTool}

Besides teams, we wanted to evaluate also with individuals, and thus we asked feedback from software professionals. This is the last group of users described in Sec.~\ref{subsec_users}. We selected technically sound individuals and experts in software development, but not necessarily in security. In particular, we wanted individuals with different roles such as CEO, CTO, consultant, architect, or system developer.
We prepared a list of probable participants and reached out to them through emails. 
Four individuals took part in the evaluation, mostly employees from eSmart Systems AS and Smart Cognition AS, both of which are software companies. 
We tried to organize a workshop to introduce the tool, but it was not possible because of their availability. However, for two of the participants, we were able to describe the tool in person, in two separate meetings. 
Thus, we sent out emails with the necessary materials to perform the assessment, i.e.:
URLs for the tool and the video tutorial presenting how to use the SCT;
Presentation slides to understand the core concept of SCM and SCT (reused from the hackathon);
Description of the task, saying that the evaluation is complete after they, at minimum, create a system, add sub-systems to this system, perform the SC assessment of each sub-system to calculate the class, and finally take the survey, asking also to provide feedback while using the tool, if they had any.

\subsubsection{Major findings}\label{subsubsec_finalHotjar}

The Hotjar data from both the hackathon and the individuals were analyzed together.
\begin{description}
\item [Heatmaps:] The heatmap of the assessment page showed that the main help menu was clicked only 0.1\% of the time. However, the user help available on each component was clicked frequently. Another highly clicked part of this page was the compute class button (5.6\%), showing that users were interested in computing the class quite often, most probably because they were repeating short cycles of changing some parameters and recomputing the class.  %
One of the least components that users interacted with was the belief and weight inputs in the assessment page, even though the help icon to explain their concept was fairly clicked.
\item [Screen recordings] showed that the majority of users used the tool as expected. They first created the account and browsed through the description and then checked the main help page. After that, they followed the instruction of creating the system and adding sub-systems. Most of the users followed a similar pattern of browsing the pages and clicking on the help icons to see the details and understand better what to select. It also showed that most users did not interact with the belief functionalities (leaving these as default).

\item[Survey:] The survey showed that the users were entirely new to such classification methodologies and took 30 to 100 minutes to apply it. 
Similarly, learning this particular tool took between 15 to 60 minutes. One of the users who had security background only used 3 minutes to learn it. It was probably because of the familiarity with security terminology, and also he had an individual workshop session with us, where we gave a presentation and a demonstration of the tool.

The tool was considered usable in the planning phase by most users, with the testing phase on second place,
according to the results from the question ``In which of the DevOps phases do you think this security classification tool (or parts of it) can be used?''.

Most of the participants found the concepts of `belief and weights' to be the most unintelligible part of the tool,
which we already observed in the heatmaps and recordings. 
Surprisingly, three of the five responses found the system definition section, where one defines the system and sub-systems, difficult.
(See \cite{shrestha2020TR} for more details.)

Three of the users considered that with a basic understanding of security, anyone could apply this method. Similarly, one of them considered that software developers could apply this methodology. However, one said that it requires the skill of security experts to apply this methodology.

Four out of five found the methodology moderately easy. However, one of the users found it difficult to apply in his system because the user considered that assessing each protection criteria is not easy without deeper knowledge of the concepts that are being evaluated. However, he considered that the methodology was easy to understand. Similarly, all the users considered it easy to find the help that they needed while using the application.
Another feedback was to provide more guidance to fill in the `belief and weights' parameters. 
\end{description}

\section{Discussions and Limitations}\label{sec_discussions}

The observations about the final version of the SC tool generally suggest that the tool is easy enough to be used by non-security experts. 
This encouraged us to release it as a public tool (see link on page~\pageref{subsec_web_v}).
The more experimental `beliefs and weights' part of the tool (which we purposely did not detail here) was considered not so easy. This only confirms the SCM research papers, who also considered this a complex feature.

In total, throughout all our stages of creating the tool, we saw the SCM applied to ca.~17 different IoT systems, done mostly by non-security experts or teams, through the use of our different prototype implementations.
These provided valuable feedback regarding the usability of the SC tool prototypes that we have been building, but can also be seen as useful proofs of the applicability of the original security classification methodology that we have worked with.

\vspace{2ex}
\noindent\emph{The principles}\label{sec_principlesResults} \, for a DevOps-ready Security Classification from Sec.~\ref{subsec_devsecops_req} have motivated our work. We have implemented the chosen methodology into a tool (following the external cognition approach), thus answering to Principle~\ref{devsecops_req_tool}; and we have worked and tested to make this tool easy to use for non-security experts (i.e., our choice of users was as such), to answer Principle~\ref{devsecops_req_easy}.
We did not strive much in the direction of Principle~\ref{devsecops_req_dynamic} because, having now a tool, one can do re-evaluations of the system by making the necessary changes in the evaluation parameters and re-running the class calculation. Since our tool can provide an API, we believe that Principle~\ref{devsecops_req_dynamic} (dynamicity) can easily be attained; however, this is more of an engineering task that is best left to a software development company to undertake. We leave this as further work, to be done by companies willing to take up our SC tool, or similar ones, into their DevSecOps tool-chains, since the adjustments and implementations are routine.

\vspace{2ex}
\noindent\emph{A general recipe}\label{sec_recipe} \, was thus discovered, for going from a research effort security classification methodology to a DevOps-ready tool. Any such endeavor, inspired by the present work, would include three main phases:
\begin{enumerate}
\item Make a step-based process out of the published security classification methodology. 
\item Test it in a low-fidelity computer-based implementation, where we have seen that the spreadsheets are very good for this purpose (especially cloud-based that also offer real-time and collaborative features). 
\item Implement the high-fidelity tool, like the web-based version that we did, where more of the process is hidden behind a natural interaction process with the tool that guides the user to the final class.
\end{enumerate}
This is something very familiar to the interaction design field \cite{preece2019interaction}, but not so familiar to the security tools developers and researchers. 
At the same time, choosing well the target group representatives to include both individuals and teams, with diverse expertise, is essential for usability testing in all three phases.

%% file: conclusion.tex
\section{Conclusion and Related work}

We have identified five principles for a security classification methodology to be DevOps-ready, i.e., ready to be used in a DevSecOps tool-chain.
Debatable as they might be, these principles are viewed as initial guidelines.
The major part of our work is concerned with exemplifying the process of taking an existing security classification methodology and working with it towards satisfying the five principles. To do this, we have created a tool that implements the chosen methodology (thus conforming to Principle~\ref{devsecops_req_tool}) and tested its usability (showing how it conforms with Principle~\ref{devsecops_req_easy}). We have detailed our process of evaluating such a tool for its usability, which involved participants from industry applying the various tool prototypes at different stages to ca.~17 IoT systems, 
during ca.~14 workshops and larger events, 
involving as test users both teams and individuals over a period of ca.~9 months.

From the process that we have detailed in both Section~\ref{sec_manual_processes} (for the manual work with the methodology) and Section~\ref{sec_tools} (for the tool prototypes), we could extract a general recipe detailed in Section~\ref{sec_recipe}. This simple guide can be applied to other `tool-ification' endeavours done for similar security methodologies. We particularly encourage such activities since we see an increased need of usable security tools and methods, demanded by the DevSecOps culture which is becoming popular in software development companies.

The tool in itself is a contribution, as it expands the user group from security experts to non-experts, and it reduces the time that was used for such evaluations before. Companies 
can now use existing internal resources (i.e., their developers or CTOs) for evaluating the security of their system. 
It is not only that more people can contribute to making the IoT products more secure, but also more people can now use a security tool to understand what it means for a product to be secured and how to achieve that.

\input{related_works}

%% file: related_works.tex
\subsection{Related Work}

We are not aware of security classification methods (or alike) that can be used within DevSecOps.
Moreover, we have no knowledge of other usability studies as the one we did here, where a security methodology (of any kind) would be transformed into a tool using an interaction design process; let alone works that also identify principles and recipes for doing such an activity, as we did.

The most relevant related works can be found among existing tools that are used to support existing security methodologies. We will evaluate these here, since other forms of related works that look at alternative classification or security evaluation methods can be found in the respective references to the security classification methodology that we have used \cite{shrestha2020methodology,shrestha2019shems,shrestha2020building}.

There are several tools \cite{maksimov2018two} to support security experts to structure their security/safety arguments based on diagrammatic notations s.a.\ the Goal Structuring Notation (GSN) \cite{spriggs2012gsn} or Toulmin's argument model \cite{toulmin2003uses}.
NOR-STA\footnote{\url{https://nor-sta.eu/en}} is an argumentation tool, based on \cite{toulmin2003uses}, to support compliance, assurance and security cases \cite{cyra2011support} using Dempster-Shafer theory for aggregation of confidence parameters (i.e., the `belief and weights' that our final SCT implements, but which we glossed over with the purpose of simplifying the presentation).
The tool is sophisticated and has many features; however, it seems limited to strict predefined requirements, thus not appealing for DevSecOps. Moreover, we have not found usability studies done for this tool, and security experts seem to be the only target group.
CertWare is an open-source Eclipse plugin from NASA \cite{barry2011certware} for development of safety, assurance and dependability cases
that seems to be superseded by AdvoCATE \cite{advocate2012,denney2018advocate}, which provides some automation support and has been applied to real systems s.a.\ unmanned aircraft.
These last two tools work similarly to NOR-STA, are aimed specifically at security experts, and we could not find usability evaluations.

For risk assessments, STRIDE 
is a popular model (and tool\footnote{\url{https://www.microsoft.com/en-us/securityengineering/sdl/threatmodeling}}) from Microsoft for threat modelling.
In the same category, CORAS is a heavy process that requires security experts and stakeholders to work together to identify threats and risks \cite{fredriksen2002coras}. CORAS comes with a tool that uses several graphical notations, and has been applied in several real systems.
ArgueSecure\footnote{\url{https://danionita.github.io/ArgueSecure/}} is a recent graphical qualitative risk assessment and security requirement elicitation framework \cite{ionita2016arguesecure,ionita2018} that is more light-weight than the above and uses an argumentation model. 
The authors have performed usability evaluations, but the tool is rather manual and meant for the security experts. Being also attack-centric, we cannot consider this tool DevOps-ready.

%% file: appendix.tex
\section{Appendix}

\vspace{4ex}

\subsection*{Screenshots from the admin-side of Hotjar showing the features that we have used and mentioned in the paper.}

\begin{figure*}[htb]
\centering
  \includegraphics[width=\textwidth]{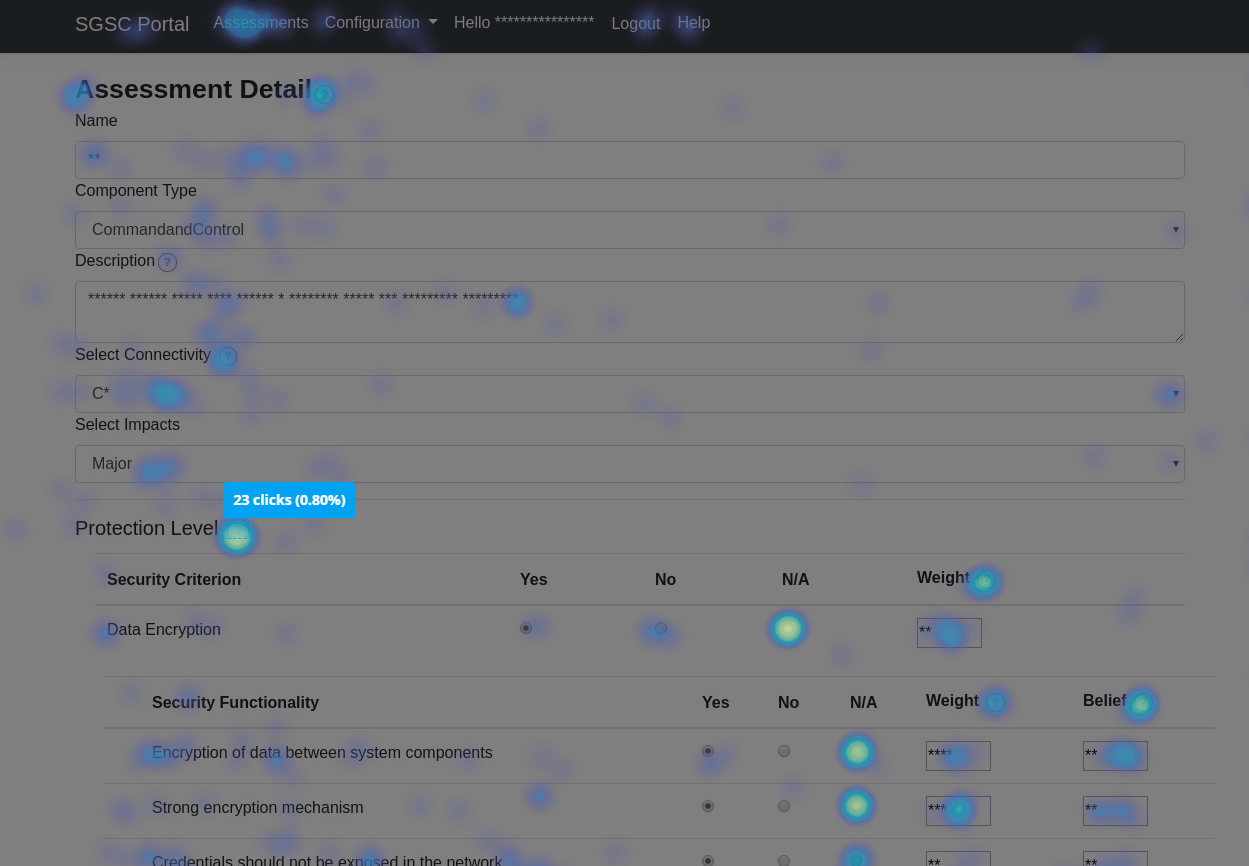}
  \caption{Snapshot taken on the admin-side of Hotjar, showing a heatmap.}
  \label{fig_heatmap}
\end{figure*}

\begin{figure*}[htb]
\centering
  \includegraphics[width=\textwidth]{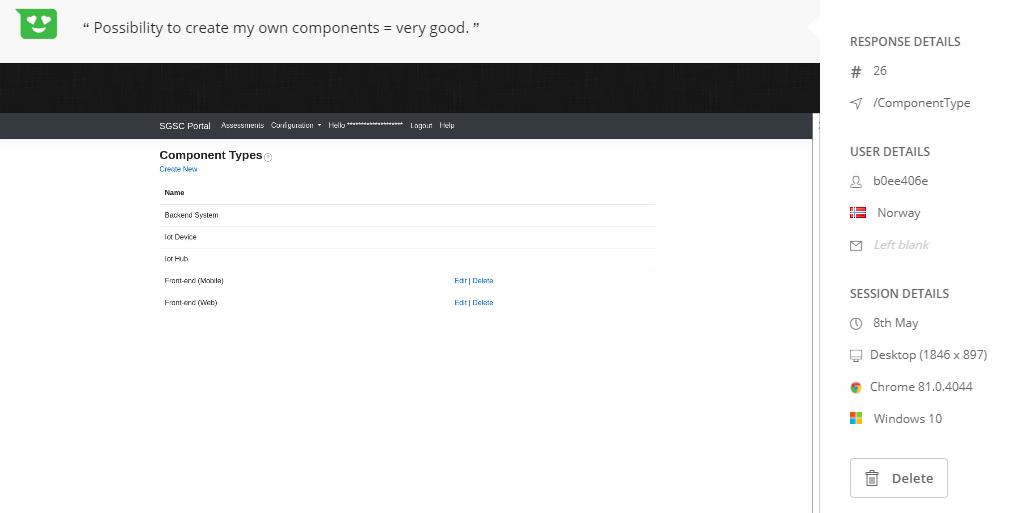}
  \caption{Snapshot taken on the admin-side of Hotjar, showing an incoming feedback.}
  \label{fig_incoming_feedback}
\end{figure*}

\begin{figure*}[htb]
\centering
  \includegraphics[width=\textwidth]{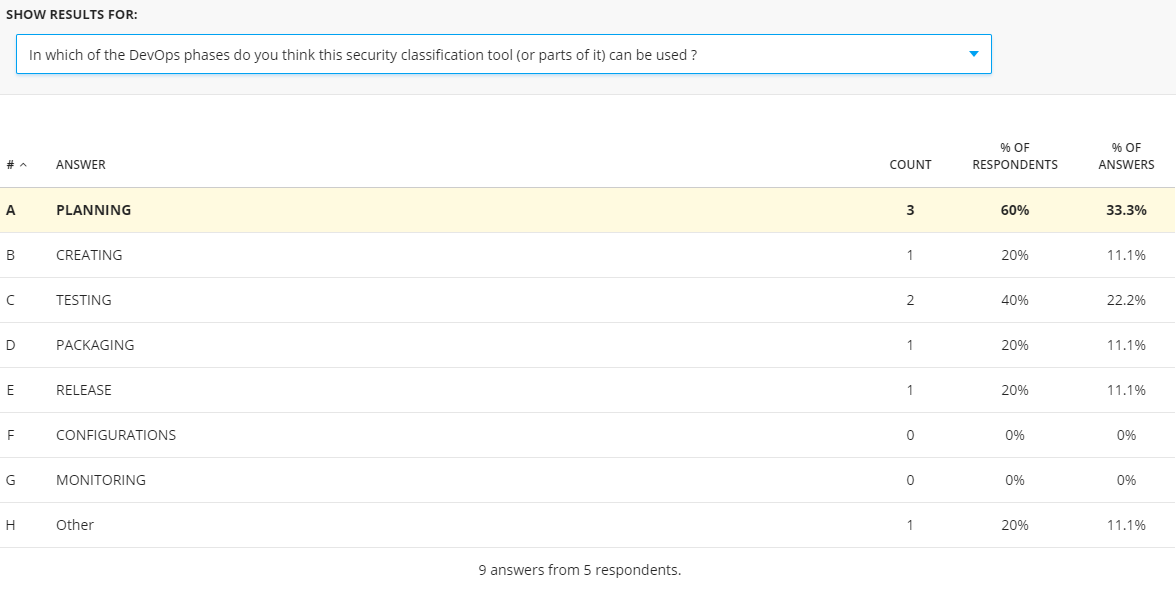}
  \caption{Survey answer on the usability of the tool in different phases.}
  \label{fig_survey_devops_phase}
\end{figure*} 

\clearpage
\newpage
\subsection*{Snapshots of all the questions from the survey about the high-fidelity prototype.}

\vspace{4ex}

\centering
  \includegraphics[width=9cm]{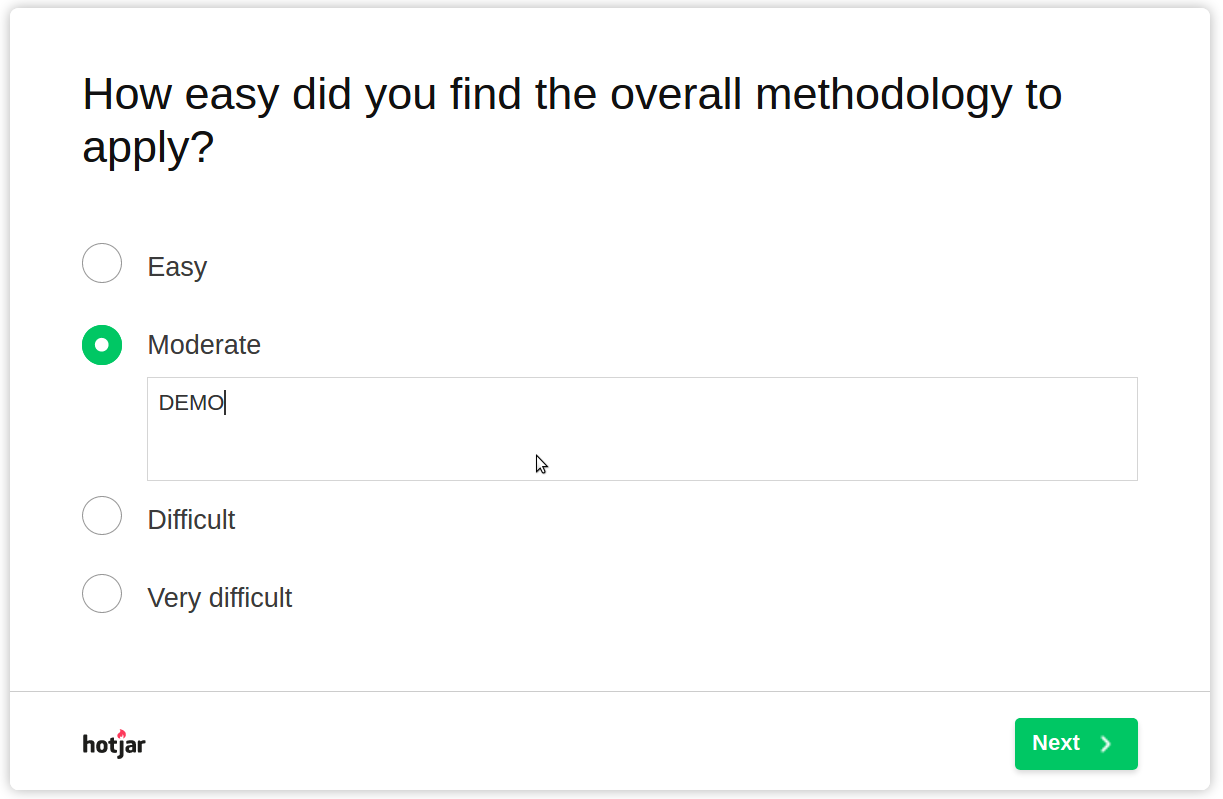}
  \includegraphics[width=9cm]{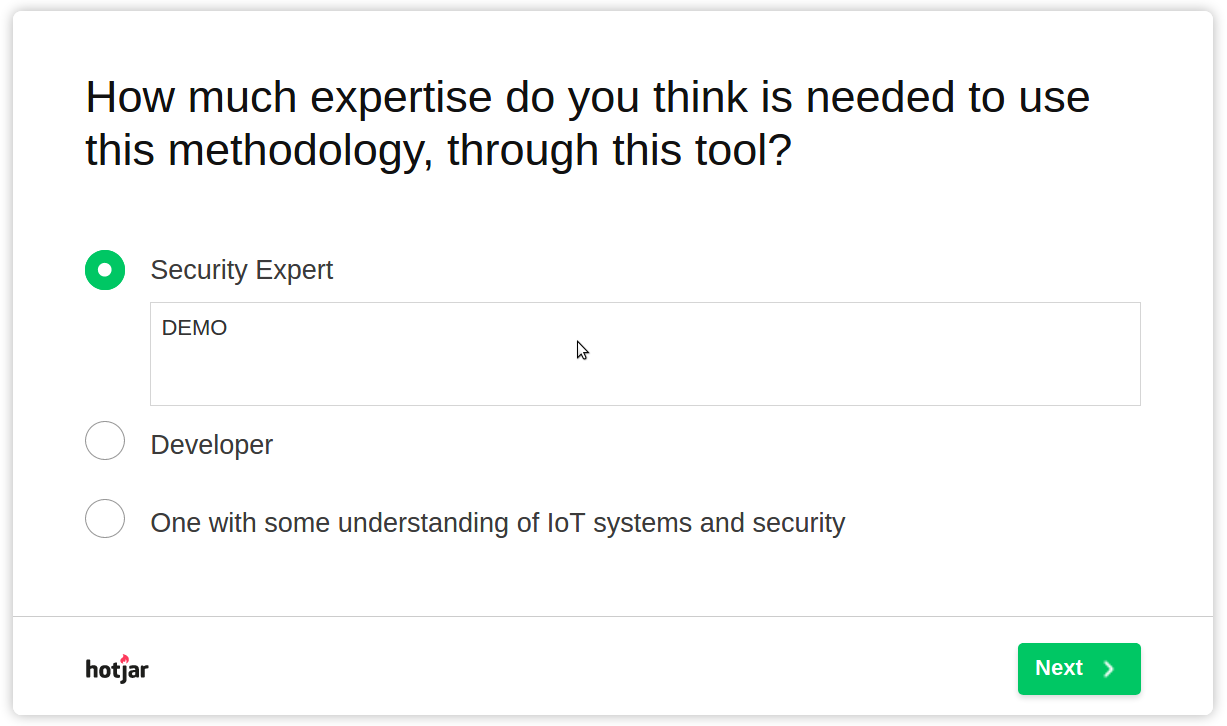}  
  \includegraphics[width=9cm]{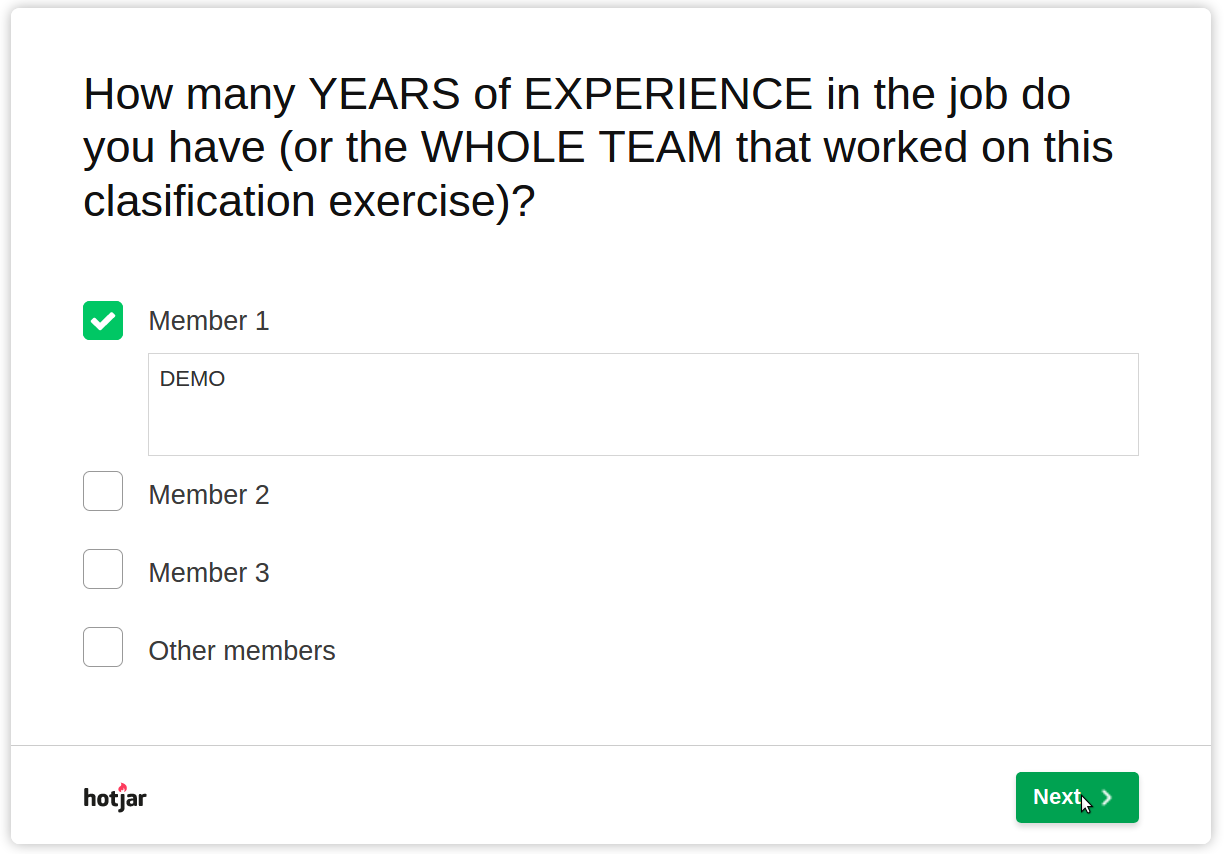}
  \includegraphics[width=9cm]{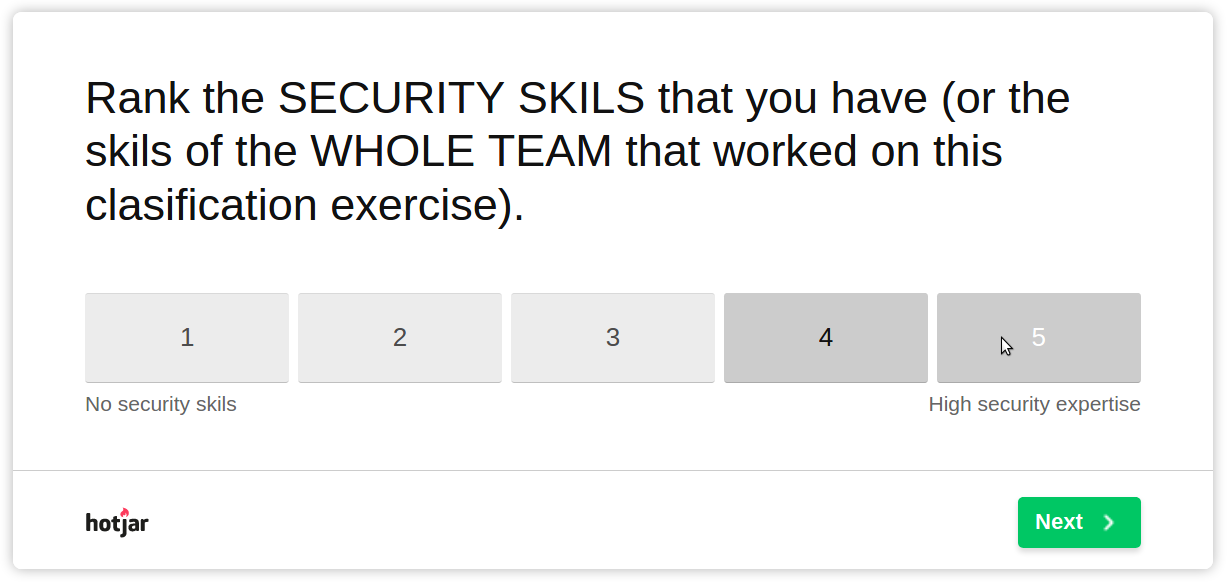}
  \includegraphics[width=9cm]{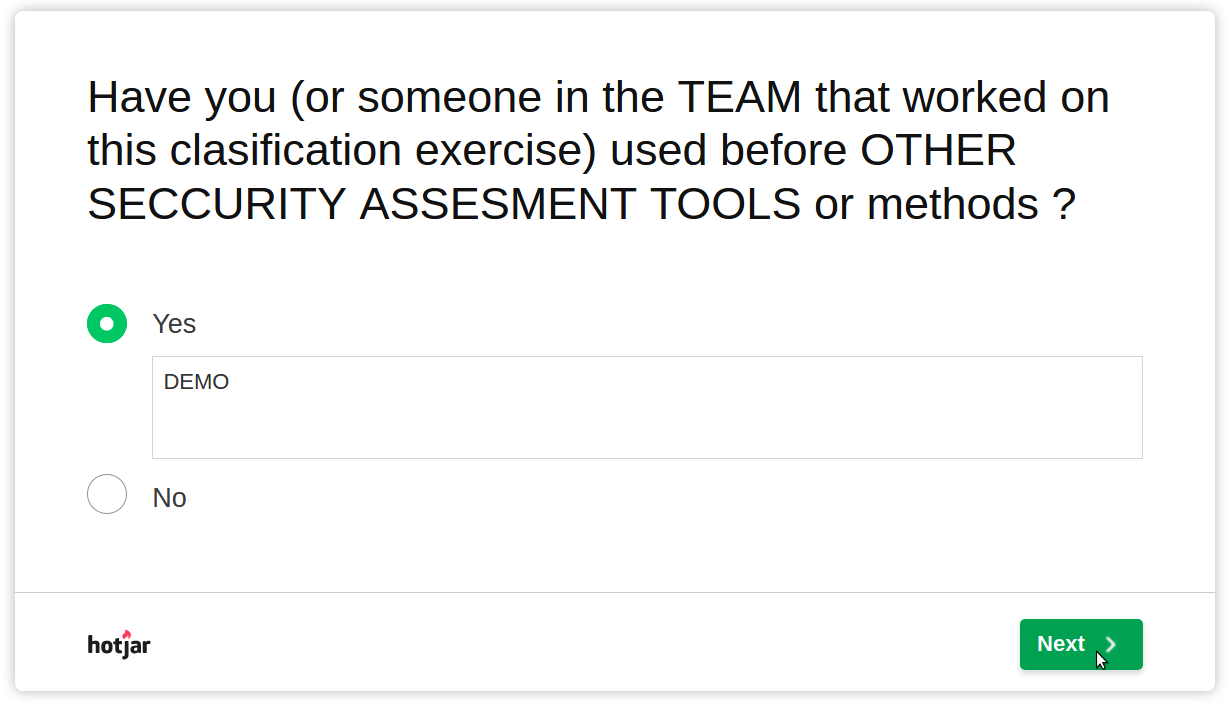}
  \includegraphics[width=9cm]{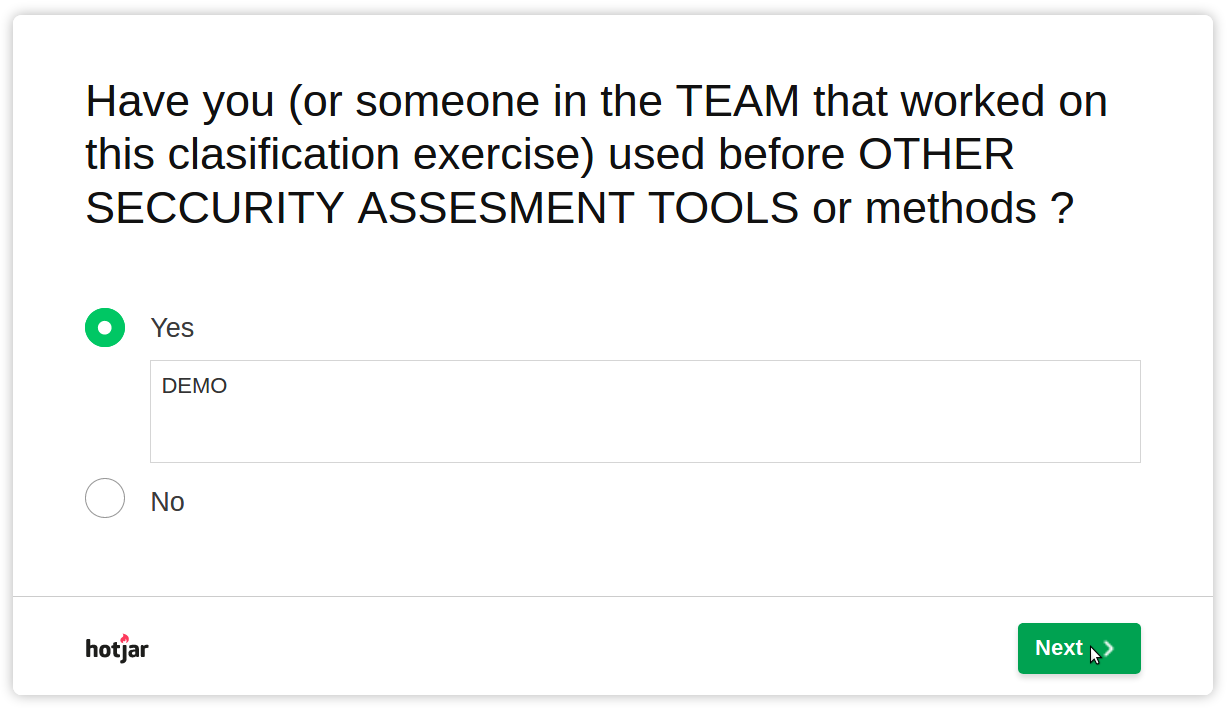}
  \includegraphics[width=9cm]{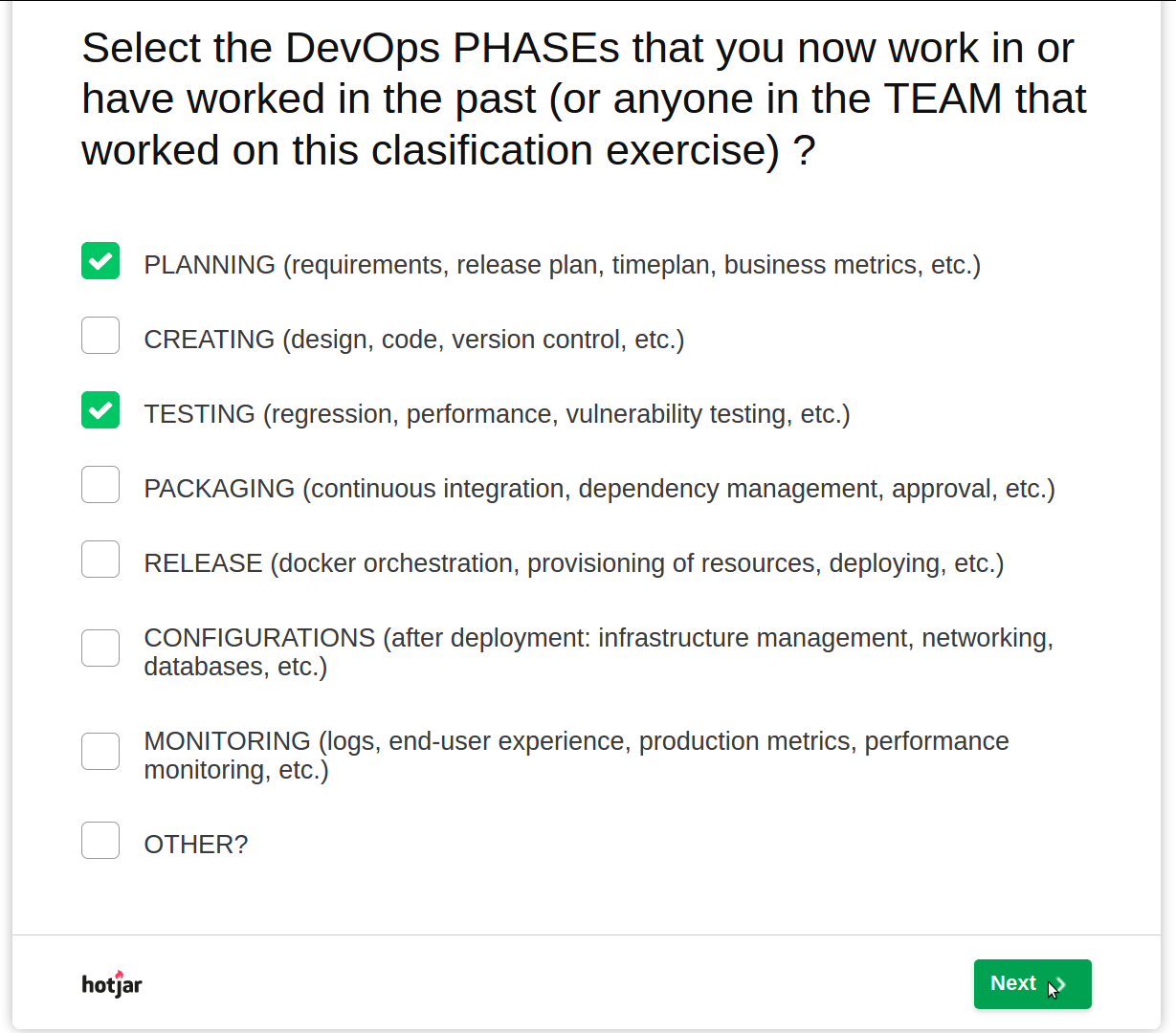}
  \includegraphics[width=9cm]{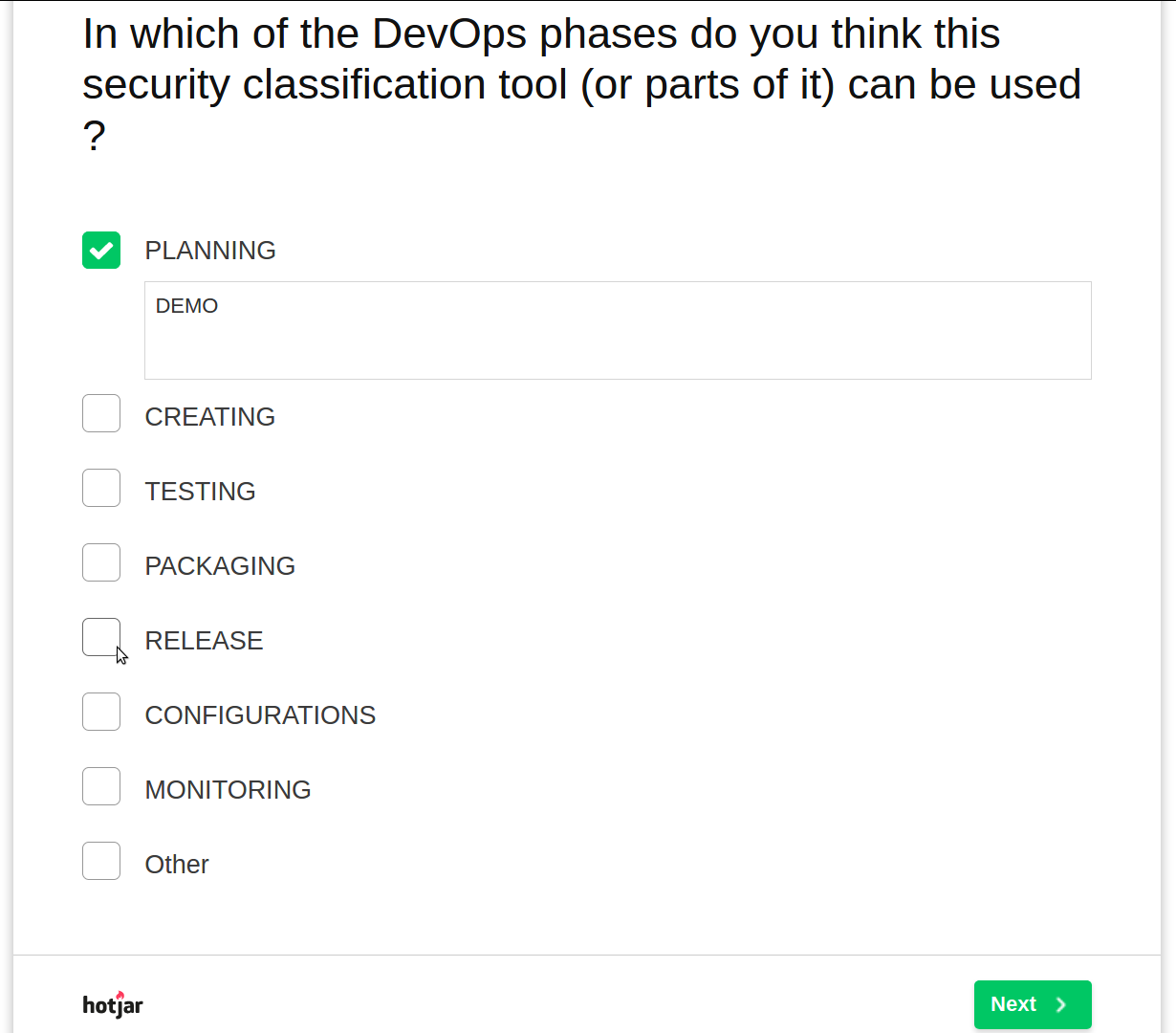}
  \includegraphics[width=9cm]{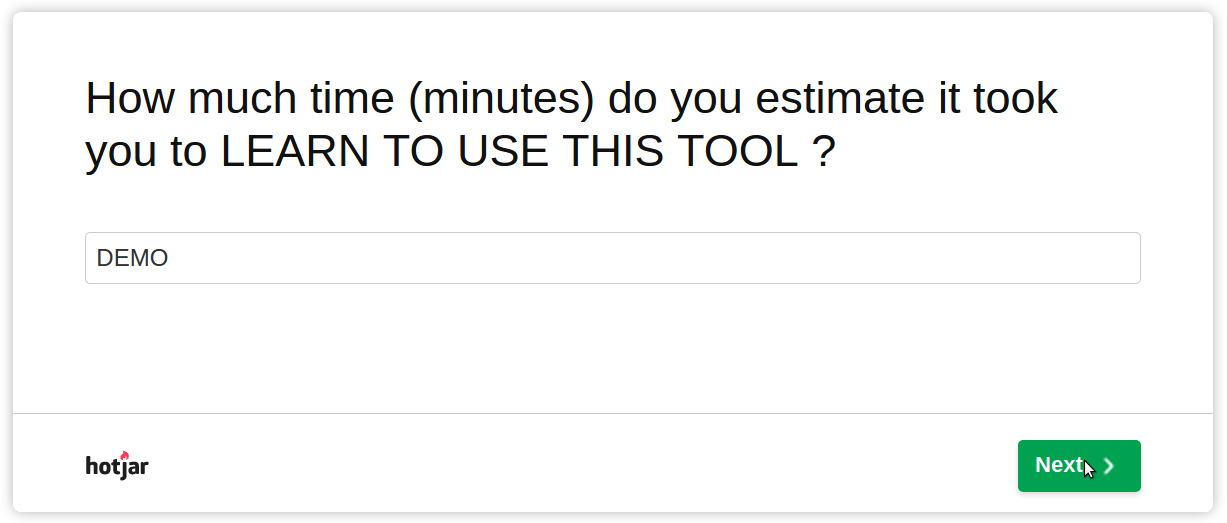}
  \includegraphics[width=9cm]{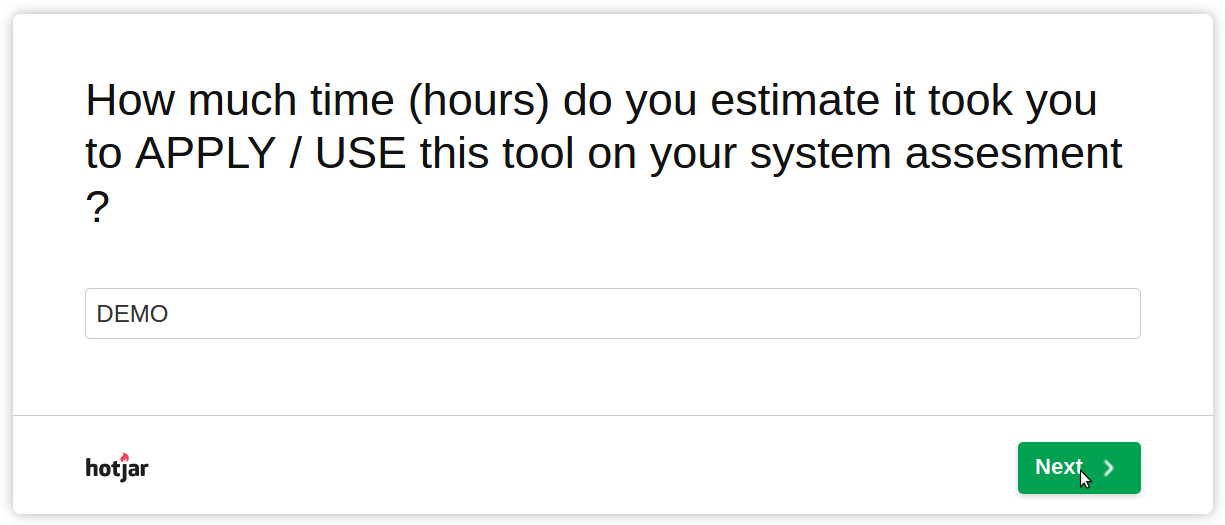}  
  \includegraphics[width=9cm]{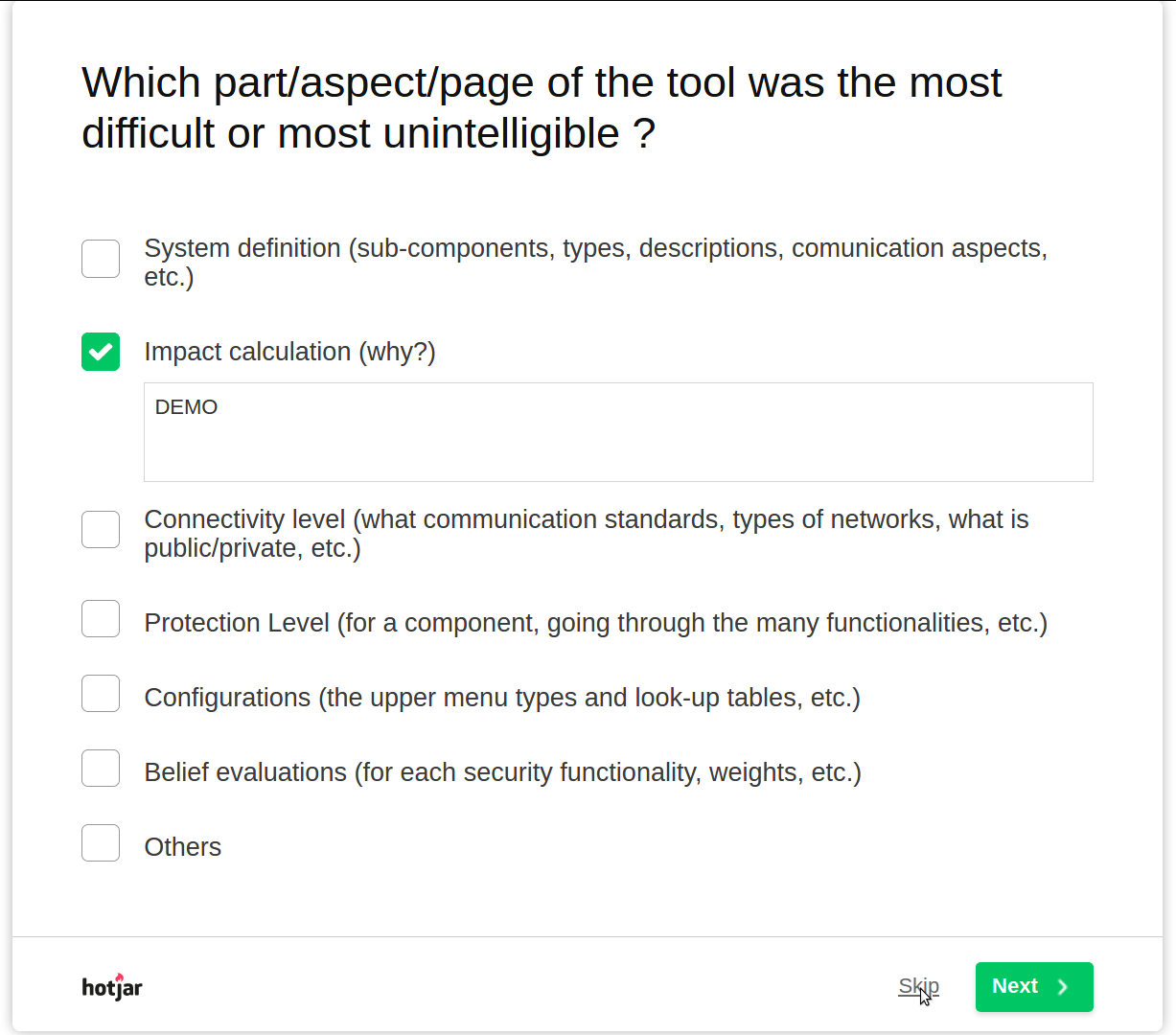}
  \includegraphics[width=9cm]{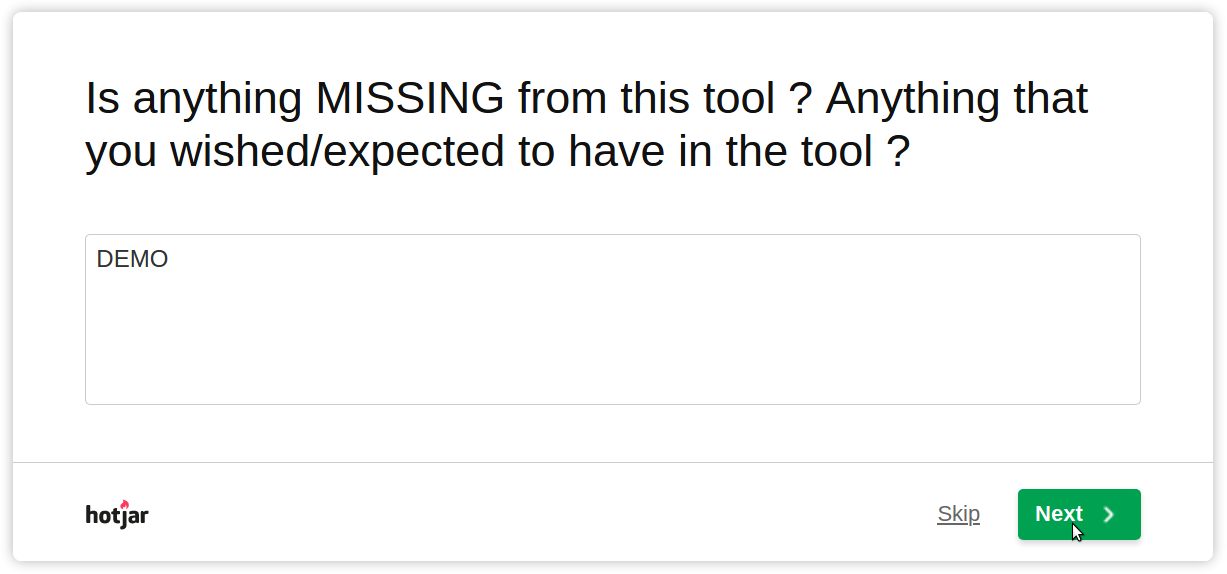}
  \includegraphics[width=9cm]{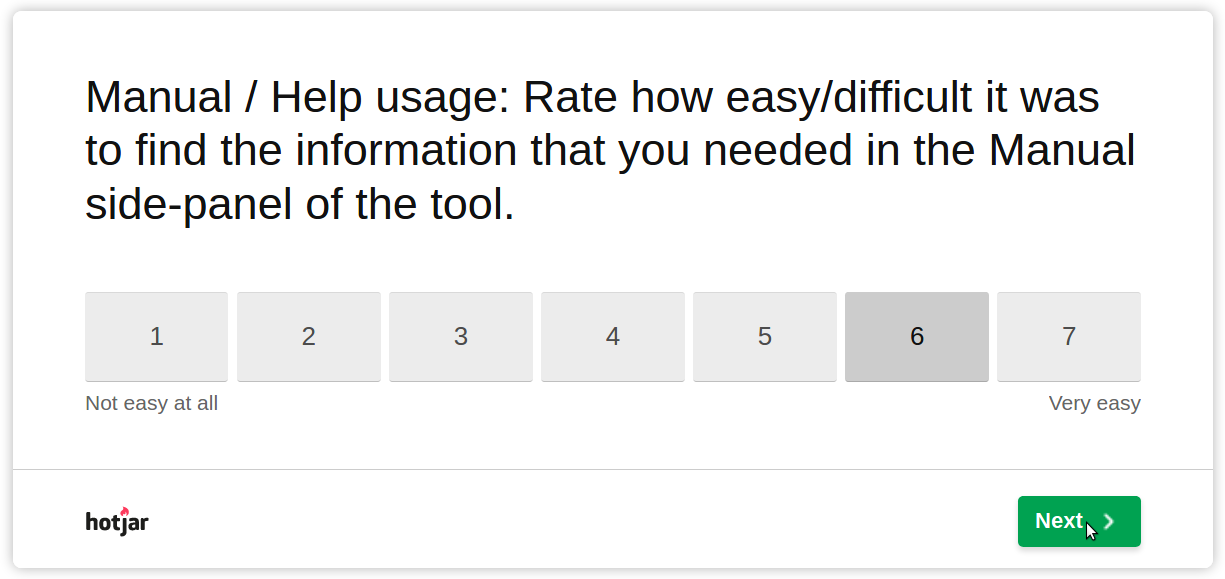}
  \includegraphics[width=9cm]{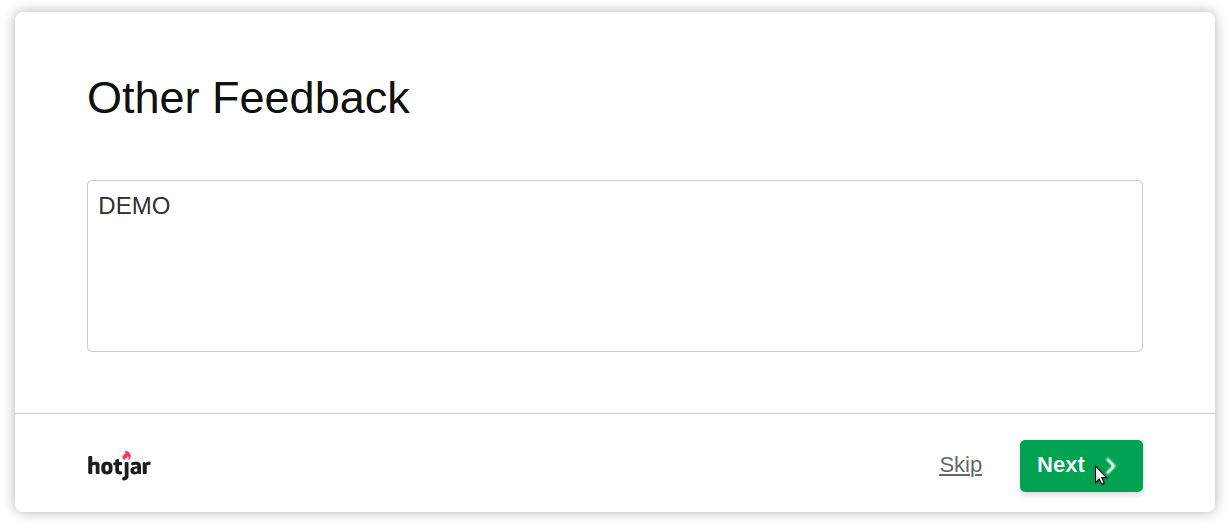}

%% file: mainArXiv2024.bbl

\begin{thebibliography}{43}


\ifx \showCODEN    \undefined \def \showCODEN     #1{\unskip}     \fi
\ifx \showDOI      \undefined \def \showDOI       #1{#1}\fi
\ifx \showISBNx    \undefined \def \showISBNx     #1{\unskip}     \fi
\ifx \showISBNxiii \undefined \def \showISBNxiii  #1{\unskip}     \fi
\ifx \showISSN     \undefined \def \showISSN      #1{\unskip}     \fi
\ifx \showLCCN     \undefined \def \showLCCN      #1{\unskip}     \fi
\ifx \shownote     \undefined \def \shownote      #1{#1}          \fi
\ifx \showarticletitle \undefined \def \showarticletitle #1{#1}   \fi
\ifx \showURL      \undefined \def \showURL       {\relax}        \fi
\providecommand\bibfield[2]{#2}
\providecommand\bibinfo[2]{#2}
\providecommand\natexlab[1]{#1}
\providecommand\showeprint[2][]{arXiv:#2}

\bibitem[\protect\citeauthoryear{Alberts, Dorofee, Stevens, and Woody}{Alberts
  et~al\mbox{.}}{2003}]%
        {alberts2003introduction}
\bibfield{author}{\bibinfo{person}{Christopher Alberts},
  \bibinfo{person}{Audrey Dorofee}, \bibinfo{person}{James Stevens}, {and}
  \bibinfo{person}{Carol Woody}.} \bibinfo{year}{2003}\natexlab{}.
\newblock \bibinfo{booktitle}{\emph{Introduction to the {OCTAVE Approach}}}.
\newblock \bibinfo{type}{{T}echnical {R}eport}.
  \bibinfo{institution}{Carnegie-Mellon University, Software Engineering
  Institute}.
\newblock


\bibitem[\protect\citeauthoryear{Anderson and Fuloria}{Anderson and
  Fuloria}{2009}]%
        {anderson2009certification}
\bibfield{author}{\bibinfo{person}{Ross Anderson} {and}
  \bibinfo{person}{Shailendra Fuloria}.} \bibinfo{year}{2009}\natexlab{}.
\newblock \showarticletitle{Certification and evaluation: A security economics
  perspective}. In \bibinfo{booktitle}{\emph{2009 IEEE Conference on Emerging
  Technologies \& Factory Automation}}. IEEE, \bibinfo{pages}{1--7}.
\newblock


\bibitem[\protect\citeauthoryear{Barry}{Barry}{2011}]%
        {barry2011certware}
\bibfield{author}{\bibinfo{person}{Matthew~R. Barry}.}
  \bibinfo{year}{2011}\natexlab{}.
\newblock \showarticletitle{CertWare: A workbench for safety case production
  and analysis}. In \bibinfo{booktitle}{\emph{2011 Aerospace conference}}.
  IEEE, \bibinfo{pages}{1--10}.
\newblock
\urldef\tempurl%
\url{https://doi.org/10.1109/AERO.2011.5747648}
\showDOI{\tempurl}


\bibitem[\protect\citeauthoryear{Boehm}{Boehm}{1988}]%
        {boehm1988spiral}
\bibfield{author}{\bibinfo{person}{Barry~W. Boehm}.}
  \bibinfo{year}{1988}\natexlab{}.
\newblock \showarticletitle{A spiral model of software development and
  enhancement}.
\newblock \bibinfo{journal}{\emph{Computer}} \bibinfo{volume}{21},
  \bibinfo{number}{5} (\bibinfo{year}{1988}), \bibinfo{pages}{61--72}.
\newblock


\bibitem[\protect\citeauthoryear{Bojanova and Voas}{Bojanova and Voas}{2017}]%
        {bojanova2017trusting}
\bibfield{author}{\bibinfo{person}{Irena Bojanova} {and}
  \bibinfo{person}{Jeffrey Voas}.} \bibinfo{year}{2017}\natexlab{}.
\newblock \showarticletitle{Trusting the Internet of Things}.
\newblock \bibinfo{journal}{\emph{IT Professional}} \bibinfo{volume}{19},
  \bibinfo{number}{5} (\bibinfo{year}{2017}), \bibinfo{pages}{16--19}.
\newblock


\bibitem[\protect\citeauthoryear{Cockburn}{Cockburn}{2006}]%
        {cockburn2006agile}
\bibfield{author}{\bibinfo{person}{Alistair Cockburn}.}
  \bibinfo{year}{2006}\natexlab{}.
\newblock \bibinfo{booktitle}{\emph{Agile software development: the cooperative
  game}}.
\newblock \bibinfo{publisher}{Pearson Education}.
\newblock


\bibitem[\protect\citeauthoryear{Cranor and Garfinkel}{Cranor and
  Garfinkel}{2005}]%
        {Cranor2005Security}
\bibfield{author}{\bibinfo{person}{Lorrie~Faith Cranor} {and}
  \bibinfo{person}{Simson Garfinkel}.} \bibinfo{year}{2005}\natexlab{}.
\newblock \bibinfo{booktitle}{\emph{{Security and usability: designing secure
  systems that people can use}}}.
\newblock \bibinfo{publisher}{{O'Reilly}}.
\newblock


\bibitem[\protect\citeauthoryear{Cyra and Gorski}{Cyra and Gorski}{2011}]%
        {cyra2011support}
\bibfield{author}{\bibinfo{person}{Lukasz Cyra} {and} \bibinfo{person}{Janusz
  Gorski}.} \bibinfo{year}{2011}\natexlab{}.
\newblock \showarticletitle{Support for argument structures review and
  assessment}.
\newblock \bibinfo{journal}{\emph{Reliability Eng. \& System Safety}}
  \bibinfo{volume}{96}, \bibinfo{number}{1} (\bibinfo{year}{2011}),
  \bibinfo{pages}{26--37}.
\newblock
\urldef\tempurl%
\url{https://doi.org/10.1016/j.ress.2010.06.027}
\showDOI{\tempurl}


\bibitem[\protect\citeauthoryear{Davis and Daniels}{Davis and Daniels}{2016}]%
        {davis2016effective}
\bibfield{author}{\bibinfo{person}{Jennifer Davis} {and} \bibinfo{person}{Ryn
  Daniels}.} \bibinfo{year}{2016}\natexlab{}.
\newblock \bibinfo{booktitle}{\emph{{Effective DevOps: building a culture of
  collaboration, affinity, and tooling at scale}}}.
\newblock \bibinfo{publisher}{O'Reilly}.
\newblock


\bibitem[\protect\citeauthoryear{Denney and Pai}{Denney and Pai}{2018}]%
        {denney2018advocate}
\bibfield{author}{\bibinfo{person}{Ewen Denney} {and} \bibinfo{person}{Ganesh
  Pai}.} \bibinfo{year}{2018}\natexlab{}.
\newblock \showarticletitle{Tool support for assurance case development}.
\newblock \bibinfo{journal}{\emph{Automated Software Engineering}}
  \bibinfo{volume}{25}, \bibinfo{number}{3} (\bibinfo{year}{2018}),
  \bibinfo{pages}{435--499}.
\newblock


\bibitem[\protect\citeauthoryear{Denney, Pai, and Pohl}{Denney
  et~al\mbox{.}}{2012}]%
        {advocate2012}
\bibfield{author}{\bibinfo{person}{Ewen Denney}, \bibinfo{person}{Ganesh Pai},
  {and} \bibinfo{person}{Josef Pohl}.} \bibinfo{year}{2012}\natexlab{}.
\newblock \showarticletitle{AdvoCATE: An Assurance Case Automation Toolset}. In
  \bibinfo{booktitle}{\emph{Computer Safety, Reliability, and Security}}
  \emph{(\bibinfo{series}{Lecture Notes in Computer Science},
  Vol.~\bibinfo{volume}{7613})}, \bibfield{editor}{\bibinfo{person}{Frank
  Ortmeier} {and} \bibinfo{person}{Peter Daniel}} (Eds.).
  \bibinfo{publisher}{Springer Berlin Heidelberg}, \bibinfo{pages}{8--21}.
\newblock
\showISBNx{978-3-642-33675-1}
\urldef\tempurl%
\url{https://doi.org/10.1007/978-3-642-33675-1_2}
\showDOI{\tempurl}


\bibitem[\protect\citeauthoryear{Dumas and Redish}{Dumas and Redish}{1999}]%
        {dumas1999practical}
\bibfield{author}{\bibinfo{person}{Joseph~S. Dumas} {and}
  \bibinfo{person}{Janice~C. Redish}.} \bibinfo{year}{1999}\natexlab{}.
\newblock \bibinfo{booktitle}{\emph{A practical guide to usability testing}}.
\newblock \bibinfo{publisher}{Intellect}.
\newblock


\bibitem[\protect\citeauthoryear{Farcic}{Farcic}{2016}]%
        {farcic2016devops}
\bibfield{author}{\bibinfo{person}{Viktor Farcic}.}
  \bibinfo{year}{2016}\natexlab{}.
\newblock \bibinfo{booktitle}{\emph{{The DevOps 2.0 Toolkit}}}.
\newblock \bibinfo{publisher}{Packt Publishing Ltd}.
\newblock


\bibitem[\protect\citeauthoryear{Forsgren, Humble, and Kim}{Forsgren
  et~al\mbox{.}}{2018}]%
        {humble2018accelerate}
\bibfield{author}{\bibinfo{person}{Nicole Forsgren}, \bibinfo{person}{Jez
  Humble}, {and} \bibinfo{person}{Gene Kim}.} \bibinfo{year}{2018}\natexlab{}.
\newblock \bibinfo{booktitle}{\emph{Accelerate: The science of lean software
  and devops: Building and scaling high performing technology organizations}}.
\newblock \bibinfo{publisher}{IT Revolution}.
\newblock


\bibitem[\protect\citeauthoryear{Franqueira, Bakalova, Tun, and
  Daneva}{Franqueira et~al\mbox{.}}{2011}]%
        {franqueira2011towards}
\bibfield{author}{\bibinfo{person}{Virginia~N.L. Franqueira},
  \bibinfo{person}{Zornitza Bakalova}, \bibinfo{person}{Thein~Than Tun}, {and}
  \bibinfo{person}{Maya Daneva}.} \bibinfo{year}{2011}\natexlab{}.
\newblock \showarticletitle{Towards agile security risk management in RE and
  beyond}. In \bibinfo{booktitle}{\emph{Workshop on Empirical Requirements
  Engineering (EmpiRE 2011)}}. IEEE, \bibinfo{pages}{33--36}.
\newblock


\bibitem[\protect\citeauthoryear{Fredriksen, Kristiansen, Gran, St{\o}len,
  Opperud, and Dimitrakos}{Fredriksen et~al\mbox{.}}{2002}]%
        {fredriksen2002coras}
\bibfield{author}{\bibinfo{person}{Rune Fredriksen}, \bibinfo{person}{Monica
  Kristiansen}, \bibinfo{person}{Bj{\o}rn~Axel Gran}, \bibinfo{person}{Ketil
  St{\o}len}, \bibinfo{person}{Tom~Arthur Opperud}, {and}
  \bibinfo{person}{Theodosis Dimitrakos}.} \bibinfo{year}{2002}\natexlab{}.
\newblock \showarticletitle{The {CORAS} framework for a model-based risk
  management process}. In \bibinfo{booktitle}{\emph{International Conference on
  Computer Safety, Reliability, and Security}},
  \bibfield{editor}{\bibinfo{person}{Stuart Anderson}, \bibinfo{person}{Massimo
  Felici}, {and} \bibinfo{person}{Sandro Bologna}} (Eds.). Springer,
  \bibinfo{pages}{94--105}.
\newblock
\urldef\tempurl%
\url{https://doi.org/10.1007/3-540-45732-1_11}
\showDOI{\tempurl}


\bibitem[\protect\citeauthoryear{Gamma, Helm, Johnson, and Vlissides}{Gamma
  et~al\mbox{.}}{1995}]%
        {gamma1994design}
\bibfield{author}{\bibinfo{person}{Erich Gamma}, \bibinfo{person}{Richard
  Helm}, \bibinfo{person}{Ralph Johnson}, {and} \bibinfo{person}{John
  Vlissides}.} \bibinfo{year}{1995}\natexlab{}.
\newblock \bibinfo{booktitle}{\emph{{Design Patterns: Elements of Reusable
  Object-Oriented Software}}}.
\newblock \bibinfo{publisher}{Addison-Wesley}.
\newblock


\bibitem[\protect\citeauthoryear{Hsu}{Hsu}{2018}]%
        {hsu2018hands}
\bibfield{author}{\bibinfo{person}{Tony Hsiang-Chih Hsu}.}
  \bibinfo{year}{2018}\natexlab{}.
\newblock \bibinfo{booktitle}{\emph{{Hands-On Security in DevOps: Ensure
  continuous security, deployment, and delivery with DevSecOps}}}.
\newblock \bibinfo{publisher}{Packt Publishing}.
\newblock


\bibitem[\protect\citeauthoryear{Humble and Farley}{Humble and Farley}{2010}]%
        {humble2010continuous}
\bibfield{author}{\bibinfo{person}{Jez Humble} {and} \bibinfo{person}{David
  Farley}.} \bibinfo{year}{2010}\natexlab{}.
\newblock \bibinfo{booktitle}{\emph{Continuous delivery: reliable software
  releases through build, test, and deployment automation}}.
\newblock \bibinfo{publisher}{Pearson Education}.
\newblock


\bibitem[\protect\citeauthoryear{Ionita, Ford, Vasenev, and Wieringa}{Ionita
  et~al\mbox{.}}{2018}]%
        {ionita2018}
\bibfield{author}{\bibinfo{person}{Dan Ionita}, \bibinfo{person}{Margaret
  Ford}, \bibinfo{person}{Alexandr Vasenev}, {and} \bibinfo{person}{Roel
  Wieringa}.} \bibinfo{year}{2018}\natexlab{}.
\newblock \showarticletitle{Graphical Modeling of Security Arguments: Current
  State and Future Directions}. In \bibinfo{booktitle}{\emph{Graphical Models
  for Security}} \emph{(\bibinfo{series}{Lecture Notes in Computer Science},
  Vol.~\bibinfo{volume}{10744})}, \bibfield{editor}{\bibinfo{person}{Peng Liu},
  \bibinfo{person}{Sjouke Mauw}, {and} \bibinfo{person}{Ketil Stolen}} (Eds.).
  \bibinfo{publisher}{Springer}, \bibinfo{pages}{1--16}.
\newblock
\showISBNx{978-3-319-74860-3}


\bibitem[\protect\citeauthoryear{Ionita, Kegel, Baltuta, and Wieringa}{Ionita
  et~al\mbox{.}}{2016}]%
        {ionita2016arguesecure}
\bibfield{author}{\bibinfo{person}{Dan Ionita}, \bibinfo{person}{Roeland
  Kegel}, \bibinfo{person}{Andrei Baltuta}, {and} \bibinfo{person}{Roel
  Wieringa}.} \bibinfo{year}{2016}\natexlab{}.
\newblock \showarticletitle{ArgueSecure: Out-of-the-box security risk
  assessment}. In \bibinfo{booktitle}{\emph{24th International Requirements
  Engineering Conference Workshops}}. IEEE, \bibinfo{pages}{74--79}.
\newblock
\urldef\tempurl%
\url{https://doi.org/10.1109/REW.2016.027}
\showDOI{\tempurl}


\bibitem[\protect\citeauthoryear{Jones}{Jones}{2004}]%
        {jones2004factor}
\bibfield{author}{\bibinfo{person}{Jack Jones}.}
  \bibinfo{year}{2004}\natexlab{}.
\newblock \bibinfo{title}{Factor analysis of information risk}.
\newblock
\newblock
\newblock
\shownote{{US Patent App. 10/912,863}.}


\bibitem[\protect\citeauthoryear{Karat, Karat, and Brodie}{Karat
  et~al\mbox{.}}{2012}]%
        {karat2012privacy}
\bibfield{author}{\bibinfo{person}{Clare-Marie Karat}, \bibinfo{person}{John
  Karat}, {and} \bibinfo{person}{Carolyn Brodie}.}
  \bibinfo{year}{2012}\natexlab{}.
\newblock \showarticletitle{{Privacy Security and Trust: Human-Computer
  Interaction Challenges and Opportunities at their Intersection}}.
\newblock In \bibinfo{booktitle}{\emph{{The Human-Computer Interaction
  Handbook: Fundamentals, Evolving Technologies, and Emerging Applications}}
  (\bibinfo{edition}{3} ed.)}, \bibfield{editor}{\bibinfo{person}{Julie~A.
  Jacko}} (Ed.). \bibinfo{publisher}{{CRC Press, Taylor \& Francis Group}},
  Chapter~29, \bibinfo{pages}{669--700}.
\newblock


\bibitem[\protect\citeauthoryear{Kennedy}{Kennedy}{1989}]%
        {kennedy1989using}
\bibfield{author}{\bibinfo{person}{Sue Kennedy}.}
  \bibinfo{year}{1989}\natexlab{}.
\newblock \showarticletitle{{Using video in the BNR usability lab}}.
\newblock \bibinfo{journal}{\emph{ACM SIGCHI Bulletin}} \bibinfo{volume}{21},
  \bibinfo{number}{2} (\bibinfo{year}{1989}), \bibinfo{pages}{92--95}.
\newblock


\bibitem[\protect\citeauthoryear{Khan and Salah}{Khan and Salah}{2018}]%
        {khan2018iot}
\bibfield{author}{\bibinfo{person}{Minhaj~Ahmad Khan} {and}
  \bibinfo{person}{Khaled Salah}.} \bibinfo{year}{2018}\natexlab{}.
\newblock \showarticletitle{{IoT security: Review, blockchain solutions, and
  open challenges}}.
\newblock \bibinfo{journal}{\emph{Future Generation Computer Systems}}
  \bibinfo{volume}{82} (\bibinfo{year}{2018}), \bibinfo{pages}{395--411}.
\newblock


\bibitem[\protect\citeauthoryear{Kim, Humble, Debois, and Willis}{Kim
  et~al\mbox{.}}{2016}]%
        {kim2016devops}
\bibfield{author}{\bibinfo{person}{Gene Kim}, \bibinfo{person}{Jez Humble},
  \bibinfo{person}{Patrick Debois}, {and} \bibinfo{person}{John Willis}.}
  \bibinfo{year}{2016}\natexlab{}.
\newblock \bibinfo{booktitle}{\emph{{The DevOps Handbook: How to Create
  World-Class Agility, Reliability, and Security in Technology
  Organizations}}}.
\newblock \bibinfo{publisher}{IT Revolution}.
\newblock


\bibitem[\protect\citeauthoryear{Lim, Ward, and Benbasat}{Lim
  et~al\mbox{.}}{1997}]%
        {lim1997empirical}
\bibfield{author}{\bibinfo{person}{Kai~H. Lim}, \bibinfo{person}{Lawrence~M.
  Ward}, {and} \bibinfo{person}{Izak Benbasat}.}
  \bibinfo{year}{1997}\natexlab{}.
\newblock \showarticletitle{An empirical study of computer system learning:
  Comparison of co-discovery and self-discovery methods}.
\newblock \bibinfo{journal}{\emph{Information Systems Research}}
  \bibinfo{volume}{8}, \bibinfo{number}{3} (\bibinfo{year}{1997}),
  \bibinfo{pages}{254--272}.
\newblock


\bibitem[\protect\citeauthoryear{{Lu} and {Xu}}{{Lu} and {Xu}}{2019}]%
        {IoT19Security}
\bibfield{author}{\bibinfo{person}{Y. {Lu}} {and} \bibinfo{person}{L.~D.
  {Xu}}.} \bibinfo{year}{2019}\natexlab{}.
\newblock \showarticletitle{{Internet of Things (IoT) Cybersecurity Research: A
  Review of Current Research Topics}}.
\newblock \bibinfo{journal}{\emph{IEEE Internet of Things Journal}}
  \bibinfo{volume}{6}, \bibinfo{number}{2} (\bibinfo{year}{2019}),
  \bibinfo{pages}{2103--2115}.
\newblock


\bibitem[\protect\citeauthoryear{Maksimov, Fung, Kokaly, and Chechik}{Maksimov
  et~al\mbox{.}}{2018}]%
        {maksimov2018two}
\bibfield{author}{\bibinfo{person}{Mike Maksimov}, \bibinfo{person}{Nick~L.S.
  Fung}, \bibinfo{person}{Sahar Kokaly}, {and} \bibinfo{person}{Marsha
  Chechik}.} \bibinfo{year}{2018}\natexlab{}.
\newblock \showarticletitle{Two decades of assurance case tools: a survey}. In
  \bibinfo{booktitle}{\emph{International Conference on Computer Safety,
  Reliability, and Security}}. Springer, \bibinfo{pages}{49--59}.
\newblock


\bibitem[\protect\citeauthoryear{Noorman, Bulck, M\"{u}hlberg, Piessens, Maene,
  Preneel, Verbauwhede, G\"{o}tzfried, M\"{u}ller, and Freiling}{Noorman
  et~al\mbox{.}}{2017}]%
        {2017IoTsecFramework}
\bibfield{author}{\bibinfo{person}{Job Noorman}, \bibinfo{person}{Jo~Van
  Bulck}, \bibinfo{person}{Jan~Tobias M\"{u}hlberg}, \bibinfo{person}{Frank
  Piessens}, \bibinfo{person}{Pieter Maene}, \bibinfo{person}{Bart Preneel},
  \bibinfo{person}{Ingrid Verbauwhede}, \bibinfo{person}{Johannes
  G\"{o}tzfried}, \bibinfo{person}{Tilo M\"{u}ller}, {and}
  \bibinfo{person}{Felix Freiling}.} \bibinfo{year}{2017}\natexlab{}.
\newblock \showarticletitle{{Sancus 2.0: A Low-Cost Security Architecture for
  IoT Devices}}.
\newblock \bibinfo{journal}{\emph{ACM Transactions on Privacy and Security
  (TOPS)}} \bibinfo{volume}{20}, \bibinfo{number}{3}, Article
  \bibinfo{articleno}{7} (\bibinfo{date}{July} \bibinfo{year}{2017}),
  \bibinfo{numpages}{33}~pages.
\newblock
\showISSN{2471-2566}
\urldef\tempurl%
\url{https://doi.org/10.1145/3079763}
\showDOI{\tempurl}


\bibitem[\protect\citeauthoryear{Nurse, Creese, and De~Roure}{Nurse
  et~al\mbox{.}}{2017}]%
        {nurse2017security}
\bibfield{author}{\bibinfo{person}{Jason~R.C. Nurse}, \bibinfo{person}{Sadie
  Creese}, {and} \bibinfo{person}{David De~Roure}.}
  \bibinfo{year}{2017}\natexlab{}.
\newblock \showarticletitle{Security risk assessment in Internet of Things
  systems}.
\newblock \bibinfo{journal}{\emph{IT professional}} \bibinfo{volume}{19},
  \bibinfo{number}{5} (\bibinfo{year}{2017}), \bibinfo{pages}{20--26}.
\newblock


\bibitem[\protect\citeauthoryear{Preece, Sharp, and Rogers}{Preece
  et~al\mbox{.}}{2019}]%
        {preece2019interaction}
\bibfield{author}{\bibinfo{person}{Jennifer Preece}, \bibinfo{person}{Helen
  Sharp}, {and} \bibinfo{person}{Yvonne Rogers}.}
  \bibinfo{year}{2019}\natexlab{}.
\newblock \bibinfo{booktitle}{\emph{Interaction design: beyond human-computer
  interaction} (\bibinfo{edition}{5} ed.)}.
\newblock \bibinfo{publisher}{John Wiley \& Sons}.
\newblock


\bibitem[\protect\citeauthoryear{Scaife and Rogers}{Scaife and Rogers}{1996}]%
        {scaife1996external}
\bibfield{author}{\bibinfo{person}{Mike Scaife} {and} \bibinfo{person}{Yvonne
  Rogers}.} \bibinfo{year}{1996}\natexlab{}.
\newblock \showarticletitle{External cognition: how do graphical
  representations work?}
\newblock \bibinfo{journal}{\emph{International journal of human-computer
  studies}} \bibinfo{volume}{45}, \bibinfo{number}{2} (\bibinfo{year}{1996}),
  \bibinfo{pages}{185--213}.
\newblock


\bibitem[\protect\citeauthoryear{Senarath, Grobler, and Arachchilage}{Senarath
  et~al\mbox{.}}{2019}]%
        {Senarath2019PEM}
\bibfield{author}{\bibinfo{person}{Awanthika Senarath},
  \bibinfo{person}{Marthie Grobler}, {and} \bibinfo{person}{Nalin
  Asanka~Gamagedara Arachchilage}.} \bibinfo{year}{2019}\natexlab{}.
\newblock \showarticletitle{{Will They Use It or Not? Investigating Software
  Developers' Intention to Follow Privacy Engineering Methodologies}}.
\newblock \bibinfo{journal}{\emph{ACM Transactions on Privacy and Security
  (TOPS)}} \bibinfo{volume}{22}, \bibinfo{number}{4}, Article
  \bibinfo{articleno}{23} (\bibinfo{date}{Nov.} \bibinfo{year}{2019}),
  \bibinfo{numpages}{30}~pages.
\newblock
\showISSN{2471-2566}
\urldef\tempurl%
\url{https://doi.org/10.1145/3364224}
\showDOI{\tempurl}


\bibitem[\protect\citeauthoryear{Shreeve, Hallett, Edwards, Anthonysamy, Frey,
  and Rashid}{Shreeve et~al\mbox{.}}{2020}]%
        {2020RiskThinking}
\bibfield{author}{\bibinfo{person}{Benjamin Shreeve}, \bibinfo{person}{Joseph
  Hallett}, \bibinfo{person}{Matthew Edwards}, \bibinfo{person}{Pauline
  Anthonysamy}, \bibinfo{person}{Sylvain Frey}, {and} \bibinfo{person}{Awais
  Rashid}.} \bibinfo{year}{2020}\natexlab{}.
\newblock \showarticletitle{{“So If Mr Blue Head Here Clicks the Link...”
  Risk Thinking in Cyber Security Decision Making}}.
\newblock \bibinfo{journal}{\emph{ACM Transactions on Privacy and Security
  (TOPS)}} \bibinfo{volume}{24}, \bibinfo{number}{1}, Article
  \bibinfo{articleno}{5} (\bibinfo{date}{Nov.} \bibinfo{year}{2020}),
  \bibinfo{numpages}{29}~pages.
\newblock
\showISSN{2471-2566}
\urldef\tempurl%
\url{https://doi.org/10.1145/3419101}
\showDOI{\tempurl}


\bibitem[\protect\citeauthoryear{Shrestha, Johansen, Moghadam, Johansen, and
  Noll}{Shrestha et~al\mbox{.}}{2020c}]%
        {shrestha2020TR}
\bibfield{author}{\bibinfo{person}{Manish Shrestha}, \bibinfo{person}{Christian
  Johansen}, \bibinfo{person}{Maunya~Doroudi Moghadam},
  \bibinfo{person}{Johanna Johansen}, {and} \bibinfo{person}{Josef Noll}.}
  \bibinfo{year}{2020}\natexlab{c}.
\newblock \bibinfo{booktitle}{\emph{Tool Support for Security Classification
  for Internet of Things (long version)}}.
\newblock \bibinfo{type}{{T}echnical {R}eport} 495.
  \bibinfo{institution}{University of Oslo}.
\newblock
\showISBNx{978-82-7368-460-8}


\bibitem[\protect\citeauthoryear{Shrestha, Johansen, and Noll}{Shrestha
  et~al\mbox{.}}{2020a}]%
        {shrestha2020building}
\bibfield{author}{\bibinfo{person}{Manish Shrestha}, \bibinfo{person}{Christian
  Johansen}, {and} \bibinfo{person}{Josef Noll}.}
  \bibinfo{year}{2020}\natexlab{a}.
\newblock \showarticletitle{{Building Confidence using Beliefs and Arguments in
  Security Class Evaluations for IoT}}. In \bibinfo{booktitle}{\emph{5th
  International Conference on Fog and Mobile Edge Computing (FMEC)}}. IEEE,
  \bibinfo{pages}{244--249}.
\newblock
\showISBNx{978-1-7281-7216-3}
\urldef\tempurl%
\url{https://doi.org/10.1109/FMEC49853.2020.9144957}
\showDOI{\tempurl}


\bibitem[\protect\citeauthoryear{Shrestha, Johansen, and Noll}{Shrestha
  et~al\mbox{.}}{2020b}]%
        {shrestha2019shems}
\bibfield{author}{\bibinfo{person}{Manish Shrestha}, \bibinfo{person}{Christian
  Johansen}, {and} \bibinfo{person}{Josef Noll}.}
  \bibinfo{year}{2020}\natexlab{b}.
\newblock \showarticletitle{{Criteria for Security Classification of Smart Home
  Energy Management Systems}}. In \bibinfo{booktitle}{\emph{Advances in Smart
  Technologies Applications and Case Studies}}. Springer.
\newblock
\urldef\tempurl%
\url{https://doi.org/10.1007/978-3-030-53187-4_19}
\showDOI{\tempurl}


\bibitem[\protect\citeauthoryear{Shrestha, Johansen, Noll, and
  Roverso}{Shrestha et~al\mbox{.}}{2020d}]%
        {shrestha2020methodology}
\bibfield{author}{\bibinfo{person}{Manish Shrestha}, \bibinfo{person}{Christian
  Johansen}, \bibinfo{person}{Josef Noll}, {and} \bibinfo{person}{Davide
  Roverso}.} \bibinfo{year}{2020}\natexlab{d}.
\newblock \showarticletitle{{A Methodology for Security Classification applied
  to Smart Grid Infrastructures}}.
\newblock \bibinfo{journal}{\emph{International Journal of Critical
  Infrastructure Protection}}  \bibinfo{volume}{28} (\bibinfo{year}{2020}),
  \bibinfo{pages}{100342}.
\newblock
\urldef\tempurl%
\url{https://doi.org/10.1016/j.ijcip.2020.100342}
\showDOI{\tempurl}


\bibitem[\protect\citeauthoryear{Spriggs}{Spriggs}{2012}]%
        {spriggs2012gsn}
\bibfield{author}{\bibinfo{person}{John Spriggs}.}
  \bibinfo{year}{2012}\natexlab{}.
\newblock \bibinfo{booktitle}{\emph{GSN-The Goal Structuring Notation: A
  Structured Approach to Presenting Arguments}}.
\newblock \bibinfo{publisher}{Springer Science \& Business Media}.
\newblock
\urldef\tempurl%
\url{https://doi.org/10.1007/978-1-4471-2312-5}
\showDOI{\tempurl}


\bibitem[\protect\citeauthoryear{Toulmin}{Toulmin}{2003}]%
        {toulmin2003uses}
\bibfield{author}{\bibinfo{person}{Stephen~E. Toulmin}.}
  \bibinfo{year}{2003}\natexlab{}.
\newblock \bibinfo{booktitle}{\emph{The uses of argument}}.
\newblock \bibinfo{publisher}{Cambridge university press}.
\newblock
\urldef\tempurl%
\url{https://doi.org/10.1017/CBO9780511840005}
\showDOI{\tempurl}


\bibitem[\protect\citeauthoryear{Wiedemann, Forsgren, Wiesche, Gewald, and
  Krcmar}{Wiedemann et~al\mbox{.}}{2019}]%
        {wiedemann2019research}
\bibfield{author}{\bibinfo{person}{Anna Wiedemann}, \bibinfo{person}{Nicole
  Forsgren}, \bibinfo{person}{Manuel Wiesche}, \bibinfo{person}{Heiko Gewald},
  {and} \bibinfo{person}{Helmut Krcmar}.} \bibinfo{year}{2019}\natexlab{}.
\newblock \showarticletitle{Research for practice: the DevOps phenomenon}.
\newblock \bibinfo{journal}{\emph{Commun. ACM}} \bibinfo{volume}{62},
  \bibinfo{number}{8} (\bibinfo{year}{2019}), \bibinfo{pages}{44--49}.
\newblock


\bibitem[\protect\citeauthoryear{Zhang, Cho, and Shieh}{Zhang
  et~al\mbox{.}}{2015}]%
        {IoTSecurity15asiaCCS}
\bibfield{author}{\bibinfo{person}{Zhi-Kai Zhang}, \bibinfo{person}{Michael
  Cheng~Yi Cho}, {and} \bibinfo{person}{Shiuhpyng Shieh}.}
  \bibinfo{year}{2015}\natexlab{}.
\newblock \showarticletitle{{Emerging Security Threats and Countermeasures in
  IoT}}. In \bibinfo{booktitle}{\emph{10th ACM Symposium on Information,
  Computer and Communications Security}} \emph{(\bibinfo{series}{ASIA CCS
  '15})}. \bibinfo{publisher}{ACM}, \bibinfo{pages}{1--6}.
\newblock
\showISBNx{9781450332453}
\urldef\tempurl%
\url{https://doi.org/10.1145/2714576.2737091}
\showDOI{\tempurl}


\end{thebibliography}
